\documentclass[reprint,amsmath,amssymb,aps,floatfix,superscriptaddress,showpacs,prl]{revtex4-1}
\usepackage{graphicx}
\usepackage{dcolumn}
\usepackage{bm}
\usepackage{color}
\usepackage{subfigure}
\usepackage{url}

\begin{document}
{

\title{Measurement of the Iron Spectrum in Cosmic Rays \\from 10 GeV/n to 2.0 TeV/n
  with the Calorimetric Electron Telescope \\on the International Space Station}

\author{O.~Adriani}
\affiliation{Department of Physics, University of Florence, Via Sansone, 1 - 50019 Sesto, Fiorentino, Italy}
\affiliation{INFN Sezione di Florence, Via Sansone, 1 - 50019 Sesto, Fiorentino, Italy}
\author{Y.~Akaike}
 \email{yakaike@aoni.waseda.jp}
\affiliation{Waseda Research Institute for Science and Engineering, Waseda University, 17 Kikuicho,  Shinjuku, Tokyo 162-0044, Japan}
\affiliation{JEM Utilization Center, Human Spaceflight Technology Directorate, Japan Aerospace Exploration Agency, 2-1-1 Sengen, Tsukuba, Ibaraki 305-8505, Japan}
\author{K.~Asano}
\affiliation{Institute for Cosmic Ray Research, The University of Tokyo, 5-1-5 Kashiwa-no-Ha, Kashiwa, Chiba 277-8582, Japan}
\author{Y.~Asaoka}
\affiliation{Institute for Cosmic Ray Research, The University of Tokyo, 5-1-5 Kashiwa-no-Ha, Kashiwa, Chiba 277-8582, Japan}
\author{E.~Berti} 
\affiliation{Department of Physics, University of Florence, Via Sansone, 1 - 50019 Sesto, Fiorentino, Italy}
\affiliation{INFN Sezione di Florence, Via Sansone, 1 - 50019 Sesto, Fiorentino, Italy}
\author{G.~Bigongiari}
\affiliation{Department of Physical Sciences, Earth and Environment, University of Siena, via Roma 56, 53100 Siena, Italy}
\affiliation{INFN Sezione di Pisa, Polo Fibonacci, Largo B. Pontecorvo, 3 - 56127 Pisa, Italy}
\author{W.R.~Binns}
\affiliation{Department of Physics and McDonnell Center for the Space Sciences, Washington University, One Brookings Drive, St. Louis, MO 63130-4899, USA}
\author{M.~Bongi}
\affiliation{Department of Physics, University of Florence, Via Sansone, 1 - 50019 Sesto, Fiorentino, Italy}
\affiliation{INFN Sezione di Florence, Via Sansone, 1 - 50019 Sesto, Fiorentino, Italy}
\author{P.~Brogi}
\affiliation{Department of Physical Sciences, Earth and Environment, University of Siena, via Roma 56, 53100 Siena, Italy}
\affiliation{INFN Sezione di Pisa, Polo Fibonacci, Largo B. Pontecorvo, 3 - 56127 Pisa, Italy}
\author{A.~Bruno}
\affiliation{Heliospheric Physics Laboratory, NASA/GSFC, Greenbelt, Maryland 20771, USA}
\author{J.H.~Buckley}
\affiliation{Department of Physics and McDonnell Center for the Space Sciences, Washington University, One Brookings Drive, St. Louis, MO 63130-4899, USA}
\author{N.~Cannady}
\affiliation{Center for Space Sciences and Technology, University of Maryland, Baltimore County, 1000 Hilltop Circle, Baltimore, Maryland 21250, USA}
\affiliation{Astroparticle Physics Laboratory, NASA/GSFC, Greenbelt, Maryland 20771, USA}
\affiliation{Center for Research and Exploration in Space Sciences and Technology, NASA/GSFC, Greenbelt, Maryland 20771, USA}
\author{G.~Castellini}
\affiliation{Institute of Applied Physics (IFAC),  National Research Council (CNR), Via Madonna del Piano, 10, 50019 Sesto, Fiorentino, Italy}
\author{C.~Checchia}
\email[]{caterina.checchia2@unisi.it}
\affiliation{Department of Physical Sciences, Earth and Environment, University of Siena, via Roma 56, 53100 Siena, Italy}
\author{M.L.~Cherry}
\affiliation{Department of Physics and Astronomy, Louisiana State University, 202 Nicholson Hall, Baton Rouge, LA 70803, USA}
\author{G.~Collazuol}
\affiliation{Department of Physics and Astronomy, University of Padova, Via Marzolo, 8, 35131 Padova, Italy}
\affiliation{INFN Sezione di Padova, Via Marzolo, 8, 35131 Padova, Italy} 
\author{K.~Ebisawa}
\affiliation{Institute of Space and Astronautical Science, Japan Aerospace Exploration Agency, 3-1-1 Yoshinodai, Chuo, Sagamihara, Kanagawa 252-5210, Japan}
\author{H.~Fuke}
\affiliation{Institute of Space and Astronautical Science, Japan Aerospace Exploration Agency, 3-1-1 Yoshinodai, Chuo, Sagamihara, Kanagawa 252-5210, Japan}
\author{S.~Gonzi}
\affiliation{Department of Physics, University of Florence, Via Sansone, 1 - 50019 Sesto, Fiorentino, Italy}
\affiliation{INFN Sezione di Florence, Via Sansone, 1 - 50019 Sesto, Fiorentino, Italy}
\author{T.G.~Guzik}
\affiliation{Department of Physics and Astronomy, Louisiana State University, 202 Nicholson Hall, Baton Rouge, LA 70803, USA}
\author{T.~Hams}
\affiliation{Center for Space Sciences and Technology, University of Maryland, Baltimore County, 1000 Hilltop Circle, Baltimore, Maryland 21250, USA}
\author{K.~Hibino}
\affiliation{Kanagawa University, 3-27-1 Rokkakubashi, Kanagawa, Yokohama, Kanagawa 221-8686, Japan}
\author{M.~Ichimura}
\affiliation{Faculty of Science and Technology, Graduate School of Science and Technology, Hirosaki University, 3, Bunkyo, Hirosaki, Aomori 036-8561, Japan}
\author{K.~Ioka}
\affiliation{Yukawa Institute for Theoretical Physics, Kyoto University, Kitashirakawa Oiwakecho, Sakyo, Kyoto 606-8502, Japan}
\author{W.~Ishizaki}
\affiliation{Institute for Cosmic Ray Research, The University of Tokyo, 5-1-5 Kashiwa-no-Ha, Kashiwa, Chiba 277-8582, Japan}
\author{M.H.~Israel}
\affiliation{Department of Physics and McDonnell Center for the Space Sciences, Washington University, One Brookings Drive, St. Louis, MO 63130-4899, USA}
\author{K.~Kasahara}
\affiliation{Department of Electronic Information Systems, Shibaura Institute of Technology, 307 Fukasaku, Minuma, Saitama 337-8570, Japan}
\author{J.~Kataoka}
\affiliation{Waseda Research Institute for Science and Engineering, Waseda University, 3-4-1 Okubo, Shinjuku, Tokyo 169-8555, Japan}
\author{R.~Kataoka}
\affiliation{National Institute of Polar Research, 10-3, Midori-cho, Tachikawa, Tokyo 190-8518, Japan}
\author{Y.~Katayose}
\affiliation{Faculty of Engineering, Division of Intelligent Systems Engineering, Yokohama National University, 79-5 Tokiwadai, Hodogaya, Yokohama 240-8501, Japan}
\author{C.~Kato}
\affiliation{Faculty of Science, Shinshu University, 3-1-1 Asahi, Matsumoto, Nagano 390-8621, Japan}
\author{N.~Kawanaka}
\affiliation{Hakubi Center, Kyoto University, Yoshida Honmachi, Sakyo-ku, Kyoto, 606-8501, Japan}
\affiliation{Department of Astronomy, Graduate School of Science, Kyoto University, Kitashirakawa Oiwake-cho, Sakyo-ku, Kyoto, 606-8502, Japan}
\author{Y.~Kawakubo}
\affiliation{Department of Physics and Astronomy, Louisiana State University, 202 Nicholson Hall, Baton Rouge, LA 70803, USA}
\author{K.~Kobayashi}
\affiliation{Waseda Research Institute for Science and Engineering, Waseda University, 17 Kikuicho,  Shinjuku, Tokyo 162-0044, Japan}
\affiliation{JEM Utilization Center, Human Spaceflight Technology Directorate, Japan Aerospace Exploration Agency, 2-1-1 Sengen, Tsukuba, Ibaraki 305-8505, Japan}
 \author{K.~Kohri} 
\affiliation{Institute of Particle and Nuclear Studies, High Energy Accelerator Research Organization, 1-1 Oho, Tsukuba, Ibaraki, 305-0801, Japan} 
\author{H.S.~Krawczynski}
\affiliation{Department of Physics and McDonnell Center for the Space Sciences, Washington University, One Brookings Drive, St. Louis, MO 63130-4899, USA}
\author{J.F.~Krizmanic}
\affiliation{Center for Space Sciences and Technology, University of Maryland, Baltimore County, 1000 Hilltop Circle, Baltimore, Maryland 21250, USA}
\affiliation{Astroparticle Physics Laboratory, NASA/GSFC, Greenbelt, Maryland 20771, USA}
\affiliation{Center for Research and Exploration in Space Sciences and Technology, NASA/GSFC, Greenbelt, Maryland 20771, USA}
\author{J.~Link}
\affiliation{Center for Space Sciences and Technology, University of Maryland, Baltimore County, 1000 Hilltop Circle, Baltimore, Maryland 21250, USA}
\affiliation{Astroparticle Physics Laboratory, NASA/GSFC, Greenbelt, Maryland 20771, USA}
\affiliation{Center for Research and Exploration in Space Sciences and Technology, NASA/GSFC, Greenbelt, Maryland 20771, USA}
\author{P.~Maestro}
\affiliation{Department of Physical Sciences, Earth and Environment, University of Siena, via Roma 56, 53100 Siena, Italy}
\affiliation{INFN Sezione di Pisa, Polo Fibonacci, Largo B. Pontecorvo, 3 - 56127 Pisa, Italy}
\author{P.S.~Marrocchesi}
\affiliation{Department of Physical Sciences, Earth and Environment, University of Siena, via Roma 56, 53100 Siena, Italy}
\affiliation{INFN Sezione di Pisa, Polo Fibonacci, Largo B. Pontecorvo, 3 - 56127 Pisa, Italy}
\author{A.M.~Messineo}
\affiliation{University of Pisa, Polo Fibonacci, Largo B. Pontecorvo, 3 - 56127 Pisa, Italy}
\affiliation{INFN Sezione di Pisa, Polo Fibonacci, Largo B. Pontecorvo, 3 - 56127 Pisa, Italy}
\author{J.W.~Mitchell}
\affiliation{Astroparticle Physics Laboratory, NASA/GSFC, Greenbelt, MD 20771, USA}
\author{S.~Miyake}
\affiliation{Department of Electrical and Electronic Systems Engineering, National Institute of Technology, Ibaraki College, 866 Nakane, Hitachinaka, Ibaraki 312-8508 Japan}
\author{A.A.~Moiseev}
\affiliation{Department of Astronomy, University of Maryland, College Park, Maryland 20742, USA}
\affiliation{Astroparticle Physics Laboratory, NASA/GSFC, Greenbelt, Maryland 20771, USA}
\affiliation{Center for Research and Exploration in Space Sciences and Technology, NASA/GSFC, Greenbelt, Maryland 20771, USA}
\author{M.~Mori}
\affiliation{Department of Physical Sciences, College of Science and Engineering, Ritsumeikan University, Shiga 525-8577, Japan}
\author{N.~Mori}
\affiliation{INFN Sezione di Florence, Via Sansone, 1 - 50019 Sesto, Fiorentino, Italy}
\author{H.M.~Motz}
\affiliation{Faculty of Science and Engineering, Global Center for Science and Engineering, Waseda University, 3-4-1 Okubo, Shinjuku, Tokyo 169-8555, Japan}
\author{K.~Munakata}
\affiliation{Faculty of Science, Shinshu University, 3-1-1 Asahi, Matsumoto, Nagano 390-8621, Japan}
\author{S.~Nakahira}
\affiliation{Institute of Space and Astronautical Science, Japan Aerospace Exploration Agency, 3-1-1 Yoshinodai, Chuo, Sagamihara, Kanagawa 252-5210, Japan}
\author{J.~Nishimura}
\affiliation{Institute of Space and Astronautical Science, Japan Aerospace Exploration Agency, 3-1-1 Yoshinodai, Chuo, Sagamihara, Kanagawa 252-5210, Japan}
\author{G.A.~de~Nolfo}
\affiliation{Heliospheric Physics Laboratory, NASA/GSFC, Greenbelt, MD 20771, USA}
\author{S.~Okuno}
\affiliation{Kanagawa University, 3-27-1 Rokkakubashi, Kanagawa, Yokohama, Kanagawa 221-8686, Japan}
\author{J.F.~Ormes}
\affiliation{Department of Physics and Astronomy, University of Denver, Physics Building, Room 211, 2112 East Wesley Ave., Denver, CO 80208-6900, USA}
\author{N.~Ospina}
\affiliation{Department of Physics and Astronomy, University of Padova, Via Marzolo, 8, 35131 Padova, Italy}\affiliation{INFN Sezione di Padova, Via Marzolo, 8, 35131 Padova, Italy} 
\author{S.~Ozawa}
\affiliation{Quantum ICT Advanced Development Center, National Institute of Information and Communications Technology, 4-2-1 Nukui-Kitamachi, Koganei, Tokyo 184-8795, Japan}
\author{L.~Pacini}
\affiliation{Department of Physics, University of Florence, Via Sansone, 1 - 50019 Sesto, Fiorentino, Italy}
\affiliation{Institute of Applied Physics (IFAC),  National Research Council (CNR), Via Madonna del Piano, 10, 50019 Sesto, Fiorentino, Italy}
\affiliation{INFN Sezione di Florence, Via Sansone, 1 - 50019 Sesto, Fiorentino, Italy}
\author{P.~Papini}
\affiliation{INFN Sezione di Florence, Via Sansone, 1 - 50019 Sesto, Fiorentino, Italy}
\author{B.F.~Rauch}
\affiliation{Department of Physics and McDonnell Center for the Space Sciences, Washington University, One Brookings Drive, St. Louis, MO 63130-4899, USA}
\author{S.B.~Ricciarini}
\affiliation{Institute of Applied Physics (IFAC),  National Research Council (CNR), Via Madonna del Piano, 10, 50019 Sesto, Fiorentino, Italy}
\affiliation{INFN Sezione di Florence, Via Sansone, 1 - 50019 Sesto, Fiorentino, Italy}
\author{K.~Sakai}
\affiliation{Center for Space Sciences and Technology, University of Maryland, Baltimore County, 1000 Hilltop Circle, Baltimore, Maryland 21250, USA}
\affiliation{Astroparticle Physics Laboratory, NASA/GSFC, Greenbelt, Maryland 20771, USA}
\affiliation{Center for Research and Exploration in Space Sciences and Technology, NASA/GSFC, Greenbelt, Maryland 20771, USA}
\author{T.~Sakamoto}
\affiliation{College of Science and Engineering, Department of Physics and Mathematics, Aoyama Gakuin University,  5-10-1 Fuchinobe, Chuo, Sagamihara, Kanagawa 252-5258, Japan}
\author{M.~Sasaki}
\affiliation{Department of Astronomy, University of Maryland, College Park, Maryland 20742, USA}
\affiliation{Astroparticle Physics Laboratory, NASA/GSFC, Greenbelt, Maryland 20771, USA}
\affiliation{Center for Research and Exploration in Space Sciences and Technology, NASA/GSFC, Greenbelt, Maryland 20771, USA}
\author{Y.~Shimizu}
\affiliation{Kanagawa University, 3-27-1 Rokkakubashi, Kanagawa, Yokohama, Kanagawa 221-8686, Japan}
\author{A.~Shiomi}
\affiliation{College of Industrial Technology, Nihon University, 1-2-1 Izumi, Narashino, Chiba 275-8575, Japan}
\author{P.~Spillantini}
\affiliation{Department of Physics, University of Florence, Via Sansone, 1 - 50019 Sesto, Fiorentino, Italy}
\author{F.~Stolzi}
\email[]{francesco.stolzi@unisi.it}
\affiliation{Department of Physical Sciences, Earth and Environment, University of Siena, via Roma 56, 53100 Siena, Italy}
\affiliation{INFN Sezione di Pisa, Polo Fibonacci, Largo B. Pontecorvo, 3 - 56127 Pisa, Italy}
\author{S.~Sugita}
\affiliation{College of Science and Engineering, Department of Physics and Mathematics, Aoyama Gakuin University,  5-10-1 Fuchinobe, Chuo, Sagamihara, Kanagawa 252-5258, Japan}
\author{A.~Sulaj} 
\affiliation{Department of Physical Sciences, Earth and Environment, University of Siena, via Roma 56, 53100 Siena, Italy}
\affiliation{INFN Sezione di Pisa, Polo Fibonacci, Largo B. Pontecorvo, 3 - 56127 Pisa, Italy}
\author{M.~Takita}
\affiliation{Institute for Cosmic Ray Research, The University of Tokyo, 5-1-5 Kashiwa-no-Ha, Kashiwa, Chiba 277-8582, Japan}
\author{T.~Tamura}
\affiliation{Kanagawa University, 3-27-1 Rokkakubashi, Kanagawa, Yokohama, Kanagawa 221-8686, Japan}
\author{T.~Terasawa}
\affiliation{RIKEN, 2-1 Hirosawa, Wako, Saitama 351-0198, Japan}
\author{S.~Torii}
\affiliation{Waseda Research Institute for Science and Engineering, Waseda University, 17 Kikuicho,  Shinjuku, Tokyo 162-0044, Japan}
\author{Y.~Tsunesada}
\affiliation{Division of Mathematics and Physics, Graduate School of Science, Osaka City University, 3-3-138 Sugimoto, Sumiyoshi, Osaka 558-8585, Japan}
\author{Y.~Uchihori}
\affiliation{National Institutes for Quantum and Radiation Science and Technology, 4-9-1 Anagawa, Inage, Chiba 263-8555, JAPAN}
\author{E.~Vannuccini}
\affiliation{INFN Sezione di Florence, Via Sansone, 1 - 50019 Sesto, Fiorentino, Italy}
\author{J.P.~Wefel}
\affiliation{Department of Physics and Astronomy, Louisiana State University, 202 Nicholson Hall, Baton Rouge, LA 70803, USA}
\author{K.~Yamaoka}
\affiliation{Nagoya University, Furo, Chikusa, Nagoya 464-8601, Japan}
\author{S.~Yanagita}
\affiliation{College of Science, Ibaraki University, 2-1-1 Bunkyo, Mito, Ibaraki 310-8512, Japan}
\author{A.~Yoshida}
\affiliation{College of Science and Engineering, Department of Physics and Mathematics, Aoyama Gakuin University,  5-10-1 Fuchinobe, Chuo, Sagamihara, Kanagawa 252-5258, Japan}
\author{K.~Yoshida}
\affiliation{Department of Electronic Information Systems, Shibaura Institute of Technology, 307 Fukasaku, Minuma, Saitama 337-8570, Japan}

\collaboration{CALET Collaboration}

\date{\today}

\begin{abstract}
  The Calorimetric Electron Telescope (CALET), in operation on the International Space Station since 2015, collected a large sample of cosmic-ray iron over a wide energy interval.
  In this Letter a measurement of the iron spectrum is presented in the range of kinetic energy per nucleon from 10 GeV$/n$ to 2.0 TeV$/n$ allowing the inclusion of iron in the list of elements studied with unprecedented precision by space-borne instruments. The measurement is based on observations carried out from January 2016 to May 2020. The CALET instrument can identify individual nuclear species via a measurement of their electric charge with a dynamic range extending far beyond iron (up to atomic number Z = 40). The energy is measured by a homogeneous calorimeter with a total equivalent thickness of 1.2 proton interaction lengths preceded by a thin (3 radiation lengths) imaging section providing tracking and energy sampling. The analysis of the data and the detailed assessment of systematic uncertainties are described and results are compared with the findings of previous experiments.
The observed differential spectrum is consistent within the errors with previous experiments. In the region from 50 GeV$ /n $ to 2 TeV$ /n $ our present data are compatible with a single power law with spectral index $ -2.60 \pm 0.03. $
\end{abstract}

\maketitle
\section{Introduction}
Direct measurements of the energy spectra of charged cosmic rays (CR) have recently achieved a level of unprecedented precision with long term observations of individual elements.
The new data, provided by magnetic spectrometers up to their maximum detectable rigidity and by space-based and balloon-borne calorimeters (as well as transition radiation and Cherenkov detectors), revealed unexpected spectral features, most notably the onset of a progressive hardening of proton and He spectra at a few hundred GeV$/n$ 
~\cite{AMS-PROTON, AMS-HE, PAMELA-PHE, CREAM1, CREAM3, DAMPE-PROTON, CALET-PROTON}
which has also been observed for heavier nuclei~\cite{AMS-CO, AMS-Li-Be-B, AMS-Ne-Mg-Si, CREAM2HARD, CALET-CO}.
The emergence of this new scenario prompted a number of theoretical interpretations~\cite{Serpico,Malkov2012,Bernard2013,Tomassetti,Drury2011,Blasi2012,Evoli2018,Evoli2019,Ohira2011,Ohira2016,Vladimirov,Ptuskin,Thoudam}
ranging from an anomalous diffusive regime near the sources (e.g.,~\cite{Tomassetti}) to the dominance of one (or more) nearby supernova remnant (SNR)  (e.g.,~\cite{Thoudam}) in the framework of specific models of confinement and gradual release from the source. In order to discriminate among different interpretations, a precision measurement of the iron spectrum is of particular interest as iron provides favourable conditions for observations, not only because of its largest relative abundance among the heavy elements, but also for a negligible contamination from spallation of higher mass elements. At the time of writing, a compilation of direct measurements of iron in space includes the satellite experiments HEAO3-C2~\cite{HEAO} (at low energy), CRN~\cite{CRN} (on Spacelab2 aboard the Challenger Space Shuttle) and NUCLEON~\cite{NUCLEON2019}.
Balloon measurements include data from Ichimura, M. et al.~\cite{pallone} (hereafter referred to as the Sanriku experiment), providing an energy estimate using Earth's magnetic field via an accurate angular measurement with nuclear emulsions, and from balloon experiments with electronic instrumentation such as ATIC, TRACER and CREAM~\cite{ATIC2,TRACER2011,TRACER2008,CREAM2}. Recently published are also data from the spectrometer AMS-02~\cite{AMS-Fe}.

CALET is a  space-based instrument ~\cite{CALET, CALET2, CALET3} optimized for the measurement
of the all-electron spectrum~\cite{CALET-ELE2017,CALET-ELE2018},
but also designed to study individual chemical elements in CR from proton to iron and above, exploring particle energies up to the PeV scale. This can be achieved thanks to its large dynamic range, adequate calorimetric depth, accurate tracking and excellent charge identification capabilities.
In the hadronic sector, CALET already provided accurate spectral measurements of protons to 10~TeV~\cite{CALET-PROTON} and of carbon and oxygen nuclei to 2.2~TeV$ /n $~\cite{CALET-CO}.

In this Letter, we present a new measurement of the iron flux from 10~GeV$/n$ to 2.0 TeV$/n$,
based on the data collected by CALET from January 1, 2016 to May 31, 2020 aboard the International Space Station (ISS).
\section{CALET Instrument}
CALET measures the particle energy with the TASC (Total AbSorption Calorimeter), a lead-tungstate homogeneous calorimeter (27 radiation lengths (r.l.), 1.2 proton interaction lengths) preceded by a thin (3 r.l.) pre-shower IMaging Calorimeter (IMC), both covering a very large dynamic range.  
Charge identification is carried out by the CHarge Detector (CHD), a two-layered hodoscope of plastic scintillator paddles placed on top of CALET.  The CHD can resolve individual elements from Z~=~1 to Z~=~40 with excellent charge resolution.  The IMC, with 16 layers of thin scintillating fibers (read out individually), provides fine-grained tracking and an independent charge measurement, via multiple samples of specific energy loss ($ dE/dx $) in each fiber,  up to the onset of saturation which occurs for ions more highly charged than silicon.
Details on the instrument layout and the trigger system can be found in the Supplemental Material (SM) of Ref.~\cite{CALET-ELE2017}.

CALET was launched on August 19, 2015 and installed on the Japanese Experiment Module Exposure Facility of the ISS. The on-orbit commissioning phase was successfully completed in the first days of October 2015.
Calibration and test of the instrument took place at the CERN-SPS during five campaigns between 2010 and 2015 with beams of electrons, protons and relativistic ions~\cite{akaike2015, bigo, niita}.
\section{Data analysis}
The sample of flight data (FD) used in the present analysis covers a period of 1613 days of CALET operation. 
The total observation live time for  the high-energy (HE) shower trigger is $T\sim3.3\times10^4$ hours, corresponding to 85.8\% of total observation time. 

A dedicated trigger mode~\cite{CALET2017,niita} allows the selection of penetrating protons and He particles for the individual on-orbit calibration of all channels.  First, raw data are corrected for gain differences among the channels,  light output non-uniformity, and any residual dependence on time and temperature. After calibration, a track is reconstructed for each event with an associated estimate of its charge and energy.

Physics processes and interactions in the apparatus are simulated by Monte Carlo (MC) techniques, based on the EPICS package~\cite{EPICS, EPICSurl} which implements the hadronic interaction model DPMJET-III~\cite{dpmjet3prl}. The instrument configuration and detector response are detailed in the simulation code which provides digitized signals from all channels. An independent analysis based on FLUKA~\cite{FLUKA,FLUKA2} is also performed to assess the systematic uncertainties.
The particle's direction and entrance point are reconstructed and fit by a tracking algorithm based on a combinatorial Kalman filter fed with the coordinates provided by the scintillating fibers in the IMC. It identifies the incident track in the presence of background hits generated by secondary radiation backscattered from the TASC~\cite{paolo2017}.
The angular resolution is $\sim{0.08}^\circ$  for Fe and the spatial resolution for the impact point on the CHD is $\sim$180 $\mu$m.

The particle's charge $ Z $ is reconstructed by measuring the ionization deposits in the CHD. The $ dE/dx $ samples are extracted from the signals of the CHD paddles traversed by the incident particle and properly corrected for path length.
Either CHD layer provides an independent $ dE/dx $ measurement.
In order to correct for the reduction of the scintillator's light yield due to the quenching effect, a ``halo'' model~\cite{GSI} has been used to fit FD samples for each nuclear species as a function of $ Z^2 $. 
The resulting curves are then used to reconstruct a charge value in either layer ($Z_{\rm CHDX}$, $Z_{\rm CHDY}$) on an event-by-event basis~\cite{CALET-CO}.
Differently from the case of lighter nuclei, an independent charge measurement with the IMC fibers is not possible for iron due to the saturation of signals occurring for $ Z \gtrsim 14$ in the upstream layers and $ Z \gtrsim22 $ in the last four layers. 
The presence of an increasing amount of backscatters from the TASC at higher energy generates additional energy deposits in the CHD that add up to the primary particle ionization signal and may induce a wrong charge identification.
This effect causes a systematic displacement of the CHDX and CHDY charge peaks to higher values (up to 0.8 charge units) with respect to the nominal charge position. Therefore it is necessary to restore the iron peak position to its nominal value, Z = 26,  by an energy dependent charge correction applied separately to the FD and the MC data. 
A charge distribution obtained by averaging $Z_{\rm CHDX}$ and $Z_{\rm CHDY}$  is shown 
in Fig.~\ref{fig:CHDAVE_100gev}.
\begin{figure}[!htb]
 	\centering
	\includegraphics[width = \hsize]{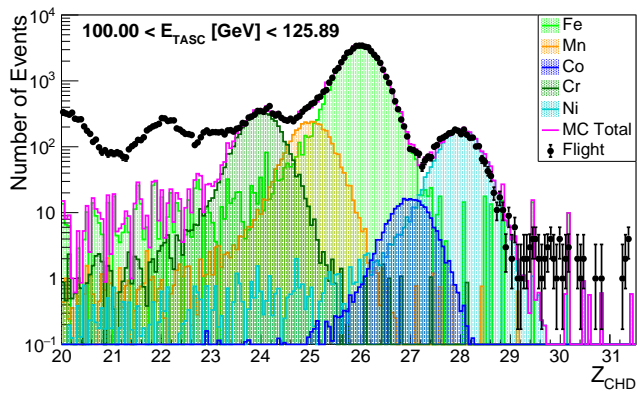}
	\caption{\scriptsize Charge distributions from the combined charge measurement of the two CHD layers in the elemental region between Ca and Ge. 
		Events are selected with $100 < E_{\rm TASC} < 125$  GeV. Flight data (black dots) are compared with Monte Carlo samples comprising chromium, manganese, iron, cobalt and nickel. Titanium and vanadium are not included in the MC because their contamination to iron data is negligible. In Fig.~S1 of the SM~\cite{PRL-SM} an enlarged version of this figure is shown, as well as the distribution for the bin $ 501 <E_{\mathrm{TASC}} < 631 $ GeV. \label{fig:CHDAVE_100gev}}
\end{figure}\noindent
The CHD charge resolution $\sigma_Z$ for iron is $\sim0.35$ (charge units).

For each event, the shower energy $E_{\rm TASC}$  is calculated as
the sum of the energy deposits of all TASC logs, after merging the gain ranges of each channel~\cite{CALET2017}.
The energy response derived from the MC simulations was tuned
using the results of a beam test carried out at the CERN-SPS in 2015~\cite{akaike2015}
with beams of accelerated ion fragments  of 13, 19 and 150~GeV$ /c/n $ momentum per nucleon (as described in the SM of Ref.~\cite{CALET-CO}).
Correction factors are 6.7\% for  $E_{\rm TASC}<45$ GeV and 3.5\% for $E_{\rm TASC}>350$~GeV, respectively. A linear interpolation is used to determine the correction factor for intermediate energies.

The onboard HE shower trigger, based on the coincidence of the summed dynode signals of the last four IMC layers and the top TASC layer (TASCX1) is fully efficient for elements heavier than oxygen.
Therefore, an offline trigger confirmation, as required for the analysis of lower charge elements~\cite{CALET-PROTON,CALET-CO}, is not necessary for iron, because the HE trigger threshold is far below the signal amplitude expected from a particle at minimum ionization (MIP) and the trigger efficiency is close to 100\%. However, in order to select interacting particles, a deposit larger than 2 sigmas of the MIP peak is required in at least one of the first four layers of the TASC.

Events with one well-fitted track crossing the whole detector from the top of the CHD to the TASC bottom layer 
(and clear from the edges of TASCX1 and of the bottom TASC layer by at least 2 cm) are selected.
The fiducial geometrical factor for this category of events is $S\Omega$~$\sim$~$416 \, \mathrm{cm}^2$sr, corresponding to about 40\% of CALET total acceptance.

Particles undergoing a charge-changing nuclear interaction in the upper part of the instrument 
are removed by requiring that the difference between the charges from either layer of the CHD is less than $1.5 $ charge units. 
Iron events are selected within an ellipse centered at Z~=~26, with 1.25~$ \sigma_x $ and 1.25~$ \sigma_y $ wide semiaxes for $ Z_{\mathrm{CHDX}} $ and $ Z_{\mathrm{CHDY}}$, respectively, and rotated clockwise by 45 degrees as shown in the cross plot of the CHDY vs CHDX charge in Fig.~S2 of the SM~\cite{PRL-SM}. 
Event selections are identical for the MC data and the FD.

For nuclei with $ Z >10 $, the TASC crystals undergo a light quenching phenomenon which is not reproduced by the MC simulations. Therefore, it is necessary to extract from the data a quenching correction to be applied
\textit{a posteriori} 
to the MC energy deposits generated in the TASC logs by non-interacting primary particles, as shown in Fig.~S3 of the SM~\cite{PRL-SM}. 

Distributions of $E_{\rm TASC}$ for Fe selected candidates are shown in Fig.~S7 of the SM~\cite{PRL-SM}, with a sample of $4.0  \times 10^4$ events.
In order to take into account the relatively li\-mi\-ted calorimetric energy resolution for hadrons (of the order of $\sim$30\%) 
energy unfolding is applied to correct for bin-to-bin migration effects. 
In this analysis, we used the Bayesian approach~\cite{Ago} 
implemented within the RooUnfold package~\cite{ROOUNFOLD} of the ROOT analysis framework~\cite{ROOT}.
Each element of the response matrix represents
the probability that a primary nucleus in a given energy interval of the CR spectrum produces 
an energy deposit falling into a given bin of $E_{\rm TASC}$.
The response matrix
is derived using the MC simulation after applying the same selection procedure as for flight data and it is shown in Fig.~S8~\cite{PRL-SM} of the SM.  

The energy spectrum is obtained from the unfolded energy distribution as follows:
\begin{equation}
\Phi(E) = \frac{N(E)}{\Delta E\;  \varepsilon(E) \;  S\Omega \;  T }
\label{eq_flux}
\end{equation}
\begin{equation}
N(E) = U \left[N_{obs}(E_{\rm TASC}) - N_{bg}(E_{\rm TASC}) \right]
\end{equation}
where $\Delta E$ denotes the energy bin width,
$E$ is the geometric mean of the lower and upper bounds of the bin~\cite{Maurino}, 
$N(E)$ the bin content in the unfolded distribution,
$\varepsilon (E)$ the total selection efficiency (Fig.~S4 of the SM~\cite{PRL-SM}), 
$U$ the unfolding procedure operator,
$N_{obs}(E_{\rm TASC})$ the bin content of observed energy distribution (including background),
and $N_{bg}(E_{\rm TASC})$ the bin content of background events in the observed energy distribution.
Background contamination from different nuclear species misidentified as Fe is shown in Fig.~S7
of the SM~\cite{PRL-SM}. 
A contamination fraction $N_{bg}/N_{obs} <1\%$ is found in the energy range between  $ 10^2 $~GeV and $ 10^3 $~GeV of $E_{\rm TASC} $ increasing up to $ \sim $2\% at $ E_{TASC} \sim 10^4 $ GeV.

\section{Systematic Uncertainties}
Dominant sources of systematics uncertainties in the iron analysis include (1) event selection,
(2) energy response,  (3) unfolding procedure and (4) MC model.
The systematic error related to charge identification (1) was studied by varying the semi-axes of the elliptical selection up to $ \pm 15\% $.
The result was an (energy bin dependent) flux variation lower than a few percent below 600 $\text{GeV}/n$ increasing to $ \sim $10\% at 1 $ \text{TeV}/n $. 

The uncertainty on the energy scale correction (2) is $\pm2$\% and depends on the accuracy of the beam test calibration. It causes a rigid shift of the measured energies, affecting the absolute flux normalization by $^{+3.3\%}_{-3.2\%}$, but not the spectral shape. As  the  beam  test  model  was  not identical  to  the  instrument  now  in  orbit,  the  difference in the spectrum obtained with either configuration was modeled and included in the systematic error.

The uncertainties due to the unfolding procedure (3) were evaluated with different response matrices computed by varying the spectral index (between -2.9 and -2.2) of the MC generation spectrum, 
or by using the Singular Value Deconvolution method, instead of the Bayesian approach, in the RooUnfold procedure~\cite{ROOUNFOLD}.

A comparison between different MC simulations (4) is in order as it is not possible to validate the MC simulations with  beam test data at high energy. A comparative study of key distributions was carried out with EPICS and FLUKA showing that the respective total selection efficiencies for Fe are in agreement within $2\%$ over the whole energy range (Figs.~S4 and S5 of the SM~\cite{PRL-SM}). However, the energy response matrices differ significantly in the low and high energy regions. The resulting fluxes show a maximum discrepancy around $10\%$ below 40 $\text{GeV}/n $, a few percent in the 100 $\text{GeV}/n $ region and less than $5 \%$ up to 1 $ \text{TeV}/n $. This turns out to be the dominant source of known systematic uncertainties at low energy.

As the trigger threshold is much smaller than the energy of a noninteracting iron event, the HE trigger efficiency is close to 100\% in the whole energy range with a negligible contribution to the systematic error. The fraction of interactions (Fig.~S6 of the SM~\cite{PRL-SM}) in the CHD, and above it, was checked by comparing the MC data and the FD as explained in the SM. The contribution due to a shower event cut, rejecting non interacting particles (5\% below 30 GeV and $ < 1 $\% above), was evaluated and included in the systematic uncertainties.   

Possible inaccuracy of track reconstruction could affect the determination of the geometrical acceptance. The contamination due to off-acceptance events that are erroneously reconstructed inside the fiducial acceptance
was estimated by MC simulation to be $\sim 1\%$ at 10 GeV$/n$ while decreasing to less than $0.1\%$ above 60 GeV$/n$. 
The systematic uncertainty on the tracking efficiency is negligible~\cite{CALET-CO}. A different tracking procedure, described in Ref.~\cite{akaike2019}, was also used to study possible systematic uncertainties in tracking efficiency. The result is well consistent with the Kalman filter algorithm. 

Additional energy-independent systematic uncertainties affecting the flux normalization include live time (3.4\%), long-term stability ($<2\%$) and geometrical factor ($ \sim 1.6\%$), as detailed in the SM of Ref.~\cite{CALET-ELE2017}. 
The flux normalization remains stable within 1\% when varying the background contamination fraction up to $ \pm 40\% $. 
The energy dependence of all systematic errors for iron analysis is shown in  Fig.~S10 of the SM~\cite{PRL-SM}. 
The total systematic error is computed as the quadrature sum of all the sources of systematics in each energy bin. 
\section{Results}
The iron differential spectrum in kinetic energy per nucleon measured by CALET from 10 GeV$/n$ to 2.0 TeV$/n$ is shown in Fig.~\ref{fig:flux}, where current uncertainties including statistical and systematic errors are bounded within a green band. The CALET spectrum is compared with the results from space-based (AMS 02~\cite{AMS-Fe}, HEAO3-C2~\cite{HEAO}, CRN~\cite{CRN}, NUCLEON~\cite{NUCLEON2019}) and balloon-borne experiments (Sanriku~\cite{pallone}, ATIC-02~\cite{ATIC2},
TRACER~\cite{TRACER2008}, CREAM-II~\cite{CREAM2}), as well as ground-based observations (H.E.S.S.~\cite{HESS}). The CALET iron flux measurements are tabulated in Table I of the SM~\cite{PRL-SM} where statistical and systematic errors are also shown. Our spectrum is consistent with ATIC 02 and TRACER at low energy and with CNR and HESS at high energy.  CALET and NUCLEON iron spectra have similar shapes while they differ in the absolute normalization of the flux. 
\begin{figure} [!htb] \centering
	\hspace*{-5mm}\includegraphics[width=1.1\hsize]{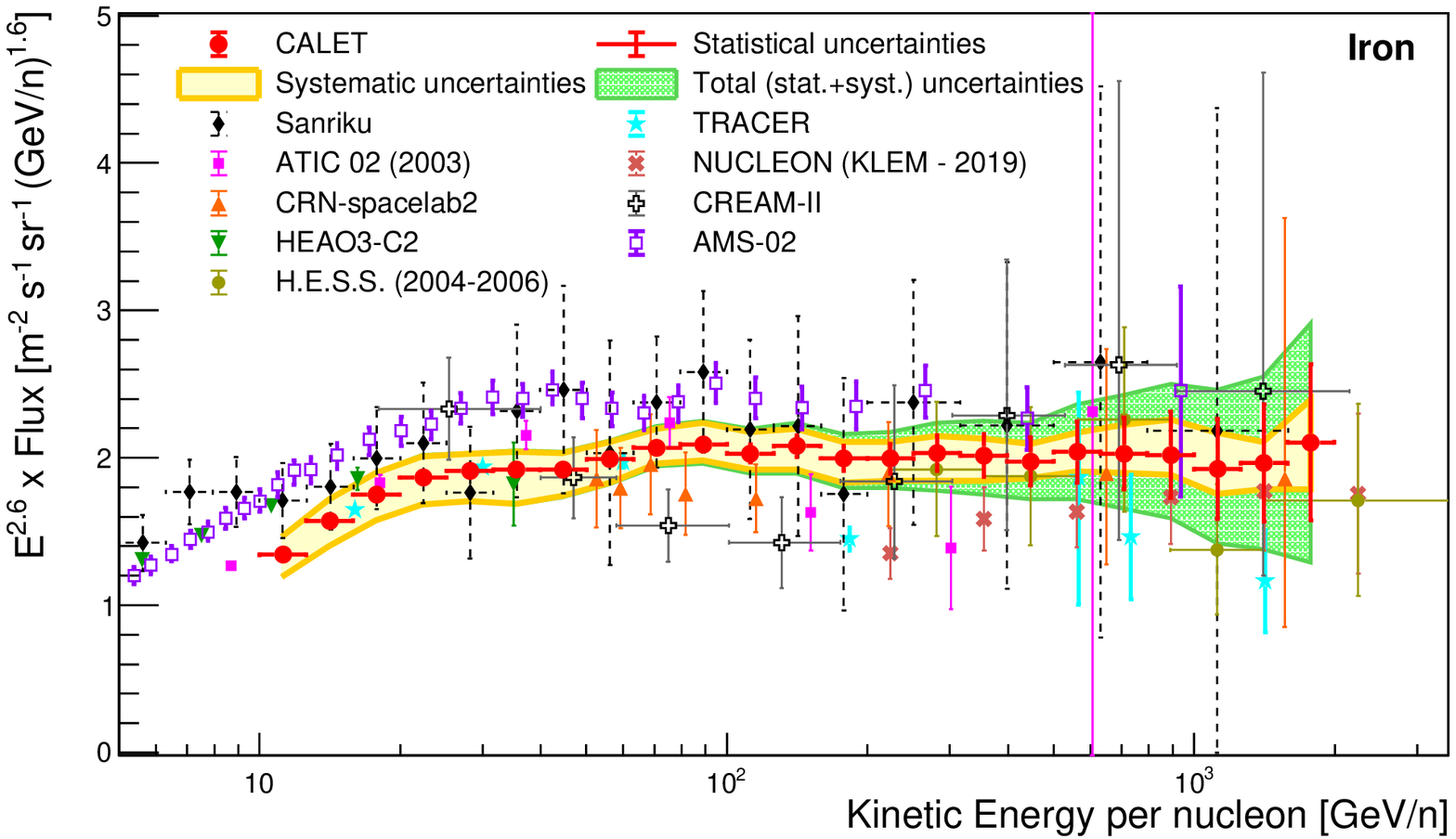}
	\caption{\scriptsize CALET iron flux (multiplied by $E^{2.6}$) as a function of kinetic energy per nucleon. Error bars of the CALET data (red) represent the statistical uncertainty only, the yellow band indicates the quadrature sum of systematic errors, while the green band indicates the quadrature sum of statistical and systematic errors. Also plotted are other direct measurements~\cite{ATIC2, TRACER2008, CREAM2, NUCLEON2019, pallone,HESS,HEAO,CRN,AMS-Fe}. This figure is reproduced enlarged in Fig.~S11 of the SM~\cite{PRL-SM}.} 
	\label{fig:flux}
\end{figure}
The latter turns out to be higher for CALET than for CRN by $\sim$10\% on average, while it is lower by 14\% with respect to Sanriku.
CALET and AMS-02 iron spectra have a very similar shape (Fig.~S12 of the SM~\cite{PRL-SM}), but differ in the absolute normalization of the flux by $ \sim 20\%$. 

Figure~\ref{fig:Fefit} shows a fit to the CALET iron flux with a single power law (SPL) function
\begin{equation}
\Phi(E) = C\, \left(\frac{E}{\text{1 GeV}/n} \right)^{\gamma}
\label{eq:SPL}
\end{equation}
where $ \gamma $ is the spectral index and $ C $ is the normalization factor. 
\begin{figure} \centering
	\includegraphics[width=\hsize]{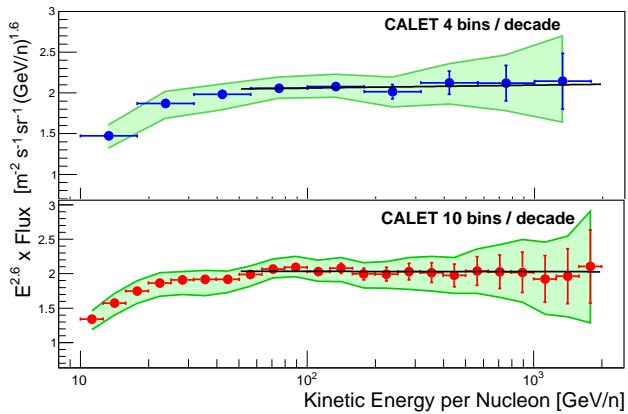}
	\caption{\scriptsize Fit of the CALET iron energy spectrum to an SPL function (black lines) in the energy range [50, 2000] GeV$ /n $ with 4 bins/decade (top) and 10 bins/decade (bottom).
		Both fluxes are multiplied  by E$^{2.6} $ where E is the kinetic energy per nucleon. The error bars are representative of purely statistical errors whereas the green band indicates the quadrature sum of statistical and systematic errors.
	}
	\label{fig:Fefit}
\end{figure}\noindent
The fit is performed from 50~GeV$/n$ to 2.0~TeV$/n$ and gives $\gamma = -2.60 \pm 0.02 (\mathrm{stat})  \pm  0.02 (\mathrm{sys})$ with $\chi^2/$d.o.f. = 4.2/14.
Furthermore, the result is stable when larger energy bins are used. As an example, when the binning is changed from 10 to 4 bins/decade (Fig.~S9 of the SM~\cite{PRL-SM}) the fit gives $ \gamma= - 2.59 \pm0.02(\mathrm{stat})\pm0.04(\mathrm{sys}) $ and the $ \chi^2/$d.o.f.~is very similar.
In order to understand whether the flux may suggest any change in spectral behavior in the region between 50~GeV$ /n $ and 2 TeV$ /n $, the spectral index $ \gamma $ is calculated by a fit of  $ d $[log($\Phi$)]/$ d $[log($ E $)]  inside a sliding window centered in each energy bin and including the neighboring~$ \pm $~3~bins. 
\begin{figure}[!htb]
	\centering
	\includegraphics[width=\hsize]{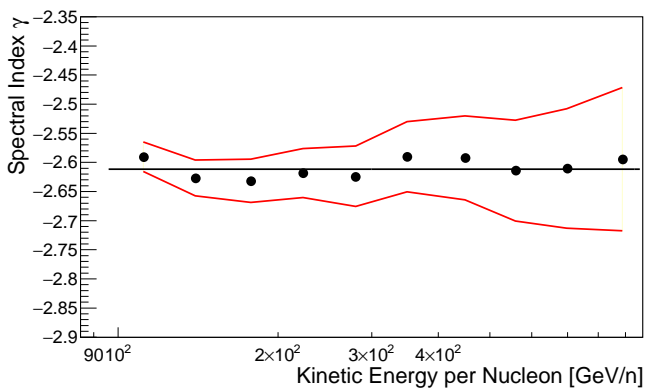}
	\caption{\scriptsize Energy dependence of the spectral index calculated within a sliding energy window for the CALET iron data.  The spectral index is determined for each bin by fitting the data using $ \pm $ 3 bins.  Red lines indicate statistical errors only. The fit with a constant function (black line) gives a mean spectral index value $ <\gamma> \, =-2.61 \pm 0.01 $.\label{fig:SlidWind}}
\end{figure}
The  result  in  Fig.~\ref{fig:SlidWind}  shows  that the iron flux, above 50 GeV$ /n $, is compatible within the errors with a single power law.

The experimental limitations of the present measurement (i.e. low statistics as well as large systematic errors for the highest energy bins) do not allow yet to test theoretical interpretations predicting spectral shapes different from a single power law. As a matter of fact, current expectations (e.g.,~\cite{Thoudam,Tomassetti}) for a detectable spectral hardening of iron are still under debate. 

\section{Conclusion}
From its privileged observation point on the ISS, CALET is carrying out direct measurements of CR fluxes extending the available spectral data on electrons and cosmic-ray nuclei to higher energies.  In this Letter (4.4 years of observations) we report a measurement of the energy spectrum of iron from 10 GeV$ /n $ to 2.0 TeV$ /n $ with a significantly better precision than most of the existing measurements. Taking into account the average size of the large systematic errors reported in the literature, our data turn out to be consistent with most of the previous measurements within the uncertainty error band, both in spectral shape and normalization. 
Below 50 GeV$ /n $ the iron spectral shape is similar to the one observed for primaries lighter than iron.
Above the same energy, our present observations are consistent with the hypothesis of an SPL spectrum up to 2 TeV$ /n $.  Beyond this limit, the uncertainties given by our present statistics and large systematics do not allow us to draw a significant conclusion on a possible deviation from a single power law. An SPL fit in this region yields a spectral index value $ \gamma = -2.60 \pm 0.03 $.
An extended data set, as expected beyond the 5 year period of continuous observations accomplished so far, will not only improve the dominant statistical limitations of the present measurement, but also our understanding of the instrument response in view of a further reduction of systematic uncertainties.

\section{Acknowledgments}
\begin{acknowledgments}
We gratefully acknowledge JAXA's contributions to the development of CALET and to the operations aboard the JEM-EF on the International Space Station.
We also wish to express our sincere gratitude to Agenzia Spaziale Italiana (ASI) and NASA for their support of the CALET project.
This work was supported in part by JSPS Grant-in-Aid for Scientific Research (S) Number 26220708 and 19H05608, 
JSPS Grant-in-Aid for Scientific Research (B) Number 17H02901, 
JSPS Grant-in-Aid for Research Activity Start-up No.20K22352 and by the
MEXT-Supported Program for the Strategic Research Foundation at Private Universities (2011-2015)
(No.S1101021) at Waseda University.
The CALET effort in Italy is supported by ASI under agreement 2013-018-R.0 and its amendments.
The CALET effort in the United States is supported by NASA through Grants No.~NNX16AB99G, No.~NNX16AC02G, and No.~NNH\-14Z\-DA0\-01N-APRA-0075. We thank Prof. G. Morlino for theoretical discussions on the iron spectral shape.
\end{acknowledgments}

\nocite{*}
\providecommand{\noopsort}[1]{}\providecommand{\singleletter}[1]{#1}%

\clearpage
	\widetext

\setcounter{equation}{0}
\setcounter{figure}{0}
\setcounter{table}{0}
\setcounter{page}{1}
\makeatletter
\renewcommand{\theequation}{S\arabic{equation}}
\renewcommand{\thefigure}{S\arabic{figure}}
\renewcommand{\bibnumfmt}[1]{[S#1]}
\renewcommand{\citenumfont}[1]{S#1}
\begin{center}
	\textbf{\large Measurement of the Iron Spectrum in Cosmic Rays \\from 10 GeV/n to 2.0 TeV/n
		with the Calorimetric Electron Telescope \\on the International Space Station
		\vspace*{0.2cm}
		SUPPLEMENTAL MATERIAL}	\\
	\vspace*{0.2cm}
	(CALET collaboration) 
\end{center}
\vspace*{1cm}
Supplemental material relative to ``Measurement of the Iron Spectrum in Cosmic Rays from 10 GeV/n to 2.0 TeV/n
with the Calorimetric Electron Telescope on the International Space Station with the Calorimetric Electron Telescope''
\vspace*{1cm}

\clearpage
\section{Charge measurement}
In Fig.~\ref{fig:Z_CHD_Feregion_SM} inclusive distributions of measured charges from flight data (FD) are compared, in two different energy bins, with Monte Carlo (MC) simulations from EPICS.
\begin{figure}[!htb] 
	\centering
	\subfigure[]
	{
		\includegraphics[scale=0.7]{Fig1.eps}  
		\label{fig:Z_CHD_Feregion_SMa}
	}
	\subfigure[]
	{
		\includegraphics[scale=0.7]{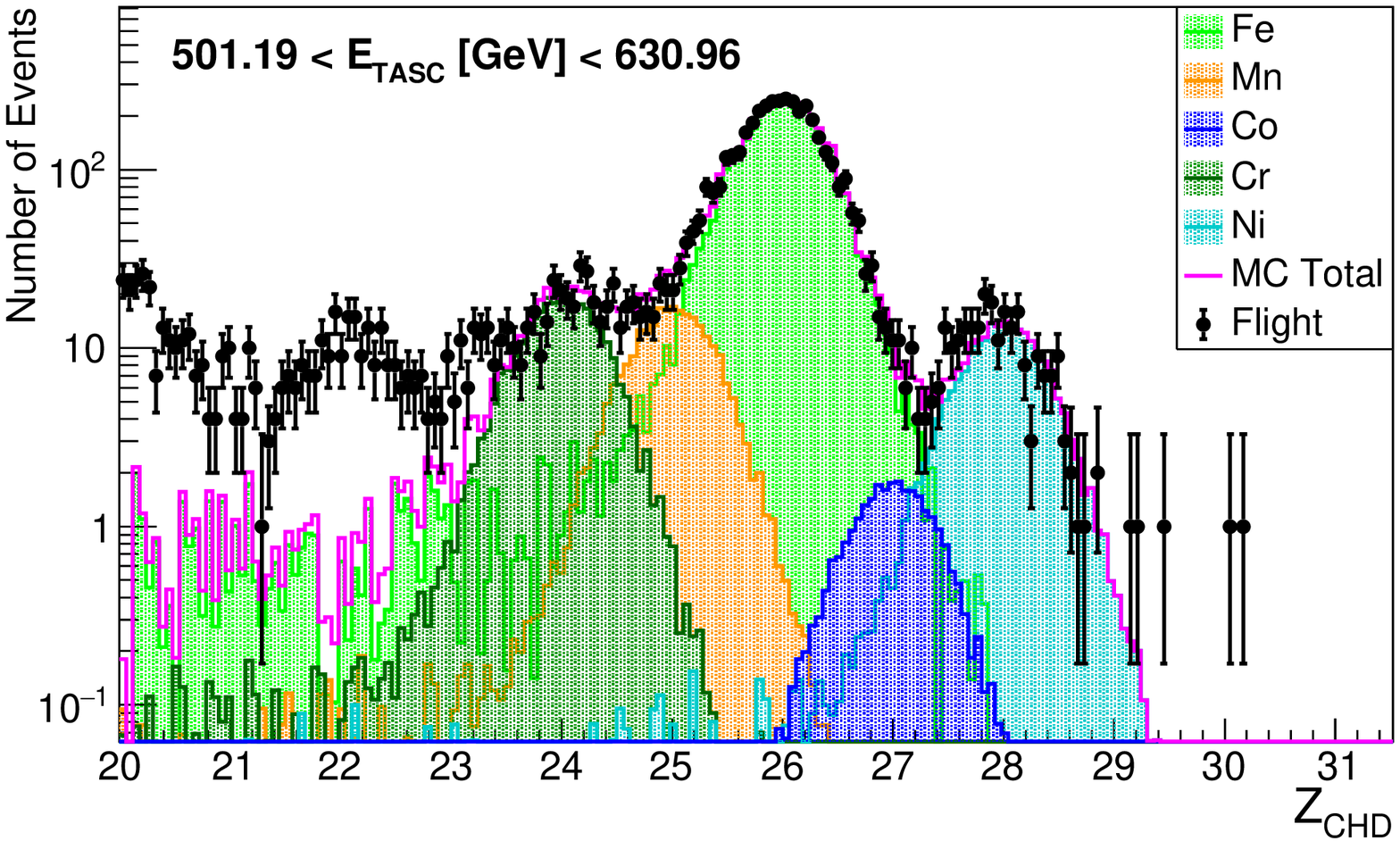}
		\label{fig:Z_CHD_Feregion_SMb}
	}
	\caption{Charge distributions from the combined charge measurement of the two CHD layers in the elemental region between Ca and Ge. 
		Events are selected with $100 < E_{\rm TASC} < 125$  GeV in (a) and $501 < E_{\rm TASC} < 630$ GeV in (b). Flight data, represented by black dots, are compared with Monte Carlo samples including chromium, manganese, iron, cobalt and nickel. Titanium and vanadium are not included in MC because their contamination to iron data is negligible.
	}
	\label{fig:Z_CHD_Feregion_SM}
\end{figure}\noindent
	
In Fig.~\ref{fig:CHDXCHDY}, obtained with flight data, a cross-plot of the independent charge measurements provided by the two CHD layers shows a clear separation of iron candidates from the less abundant neighbor elements. 

\begin{figure}[h!]  \centering
	\includegraphics[scale=0.5]{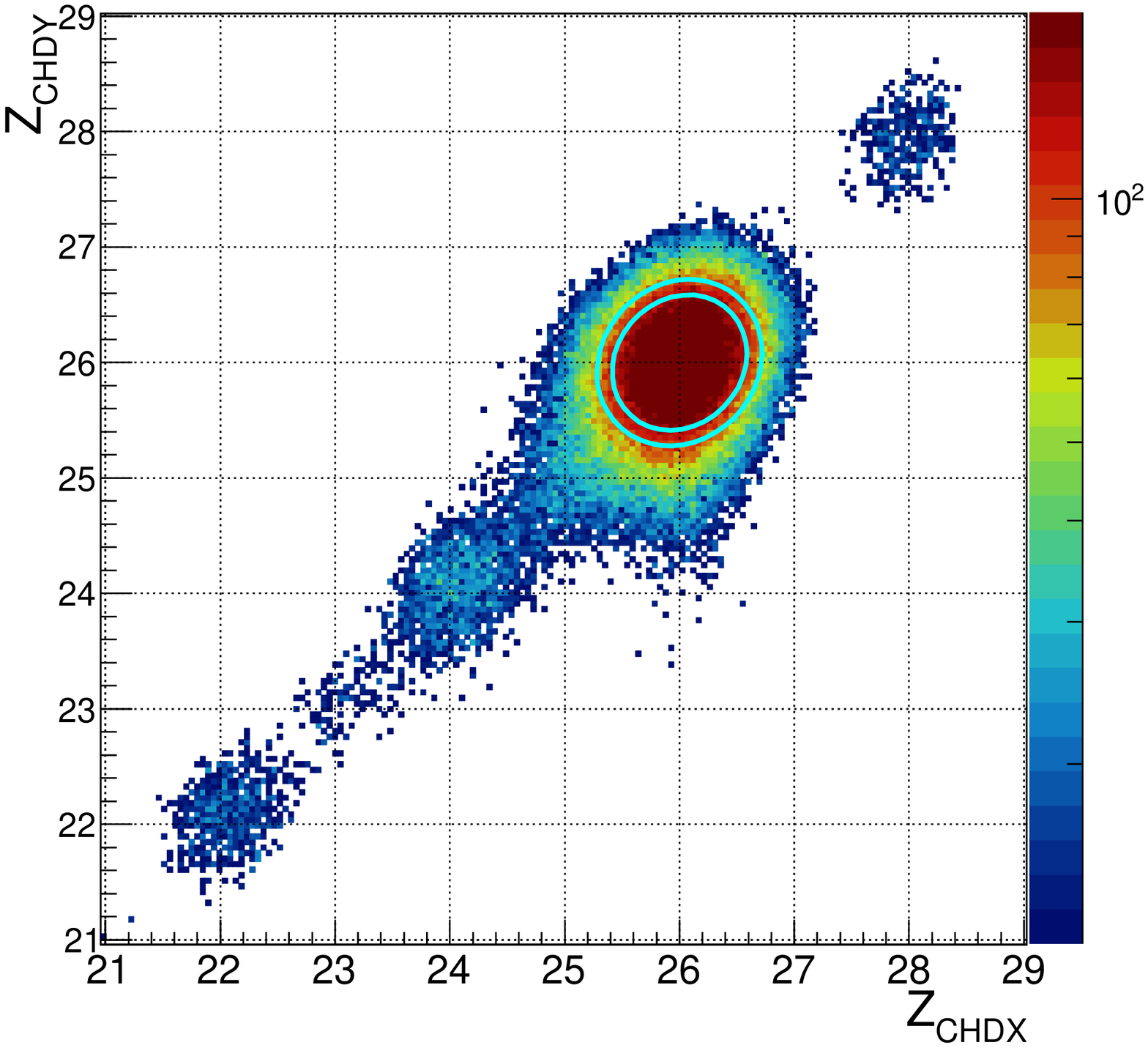}
	\caption{Crossplot of $ \mathrm{Z_{CHDY}} $ vs. $ \mathrm{Z_{CHDX}} $ reconstructed charges in the elemental range between Ti (Z~=~22) and Ni (Z~=~28). 
		Iron candidates are selected inside an ellipse with minor and major semi-axes 1.25 $ \sigma_x $ and 1.25 $\sigma_y$, respectively,  rotated clockwise by 45$^\circ $. The maximum and the minimum elliptical selection are indicated by the cyan ellipses in the figure. }
	\label{fig:CHDXCHDY}
\end{figure}\noindent

\section{Light quenching in the TASC crystals}
For nuclei with $ Z >10 $, the TASC crystals undergo a light quenching phenomenon as is clearly visible in Fig.~\ref{fig:TASCTARLE}. In the same figure the position of the peak for a minimum-ionizing particle (MIP), generated by a non-interacting primary particle crossing the first TASC layer, is plotted as a function of $ Z^2 $ for nine elements ranging from O to Ni and selected from flight data. A ``halo'' model is used in the fit to parameterize the non-linearity of the scintillator's response due to light quenching. The applied corrections on the signals from the plastic scintillators are based on the same model, as explained in~\cite{GSI-SM}.

\begin{figure}[!htb]
	\centering
	\includegraphics[width=0.5\hsize]{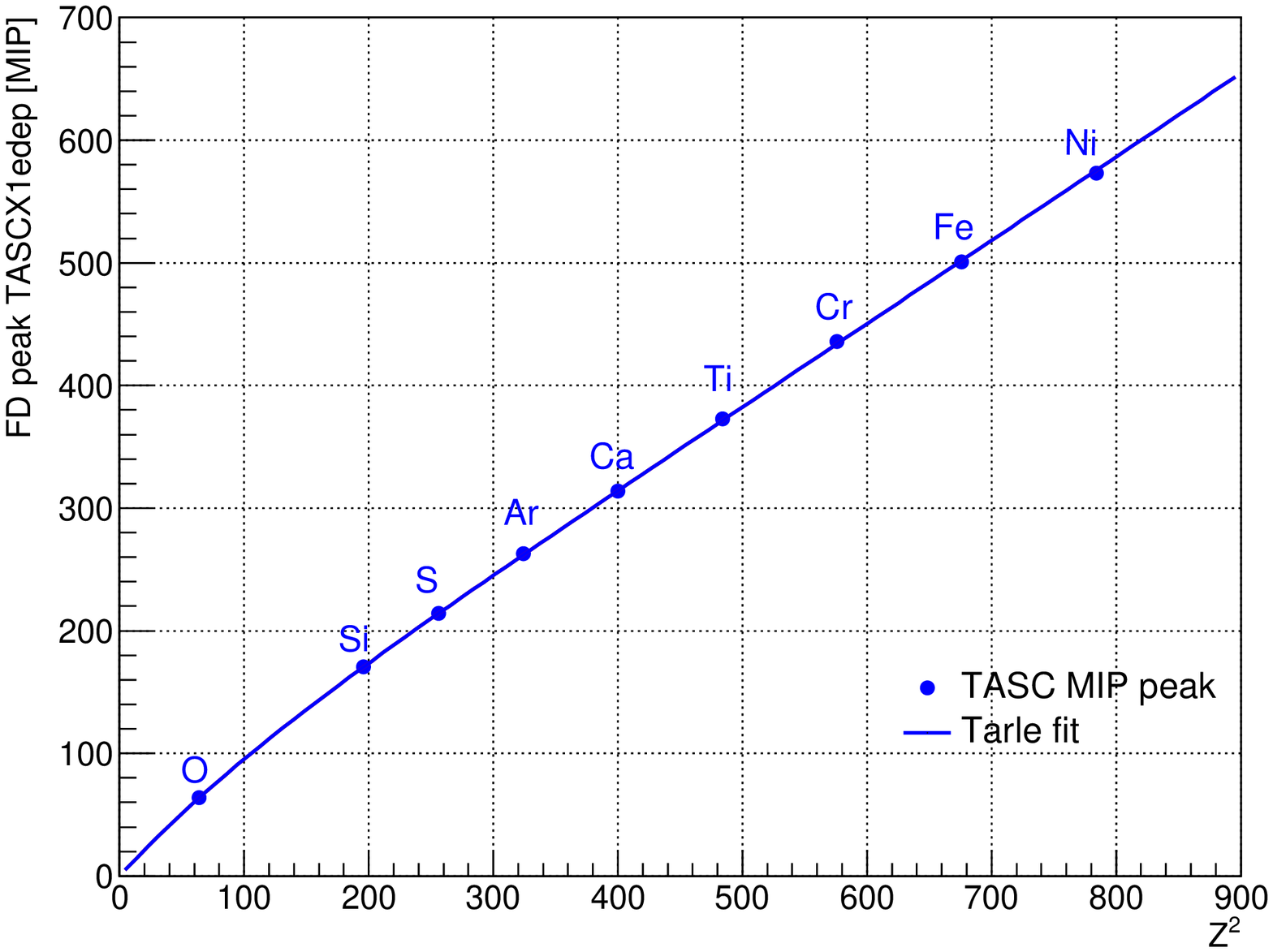}
	\caption{Cross-plot of the minimum ionizing energy deposit in the first TASC layer as a function of $ Z^2 $ for nine elements ranging from O to Ni selected from flight data.  Each element used in the fit is identified by its chemical label.\label{fig:TASCTARLE}}
\end{figure}

\clearpage
	\section{Additional information on the analysis}

\noindent{\bf{Efficiencies.}} The total efficiency and relative efficiencies (i.e., the efficiency of a given cut normalized to the previous cut) were studied extensively over the whole energy range covered by the iron flux measurement. 
The total selection efficiency from EPICS (blue open circles) and FLUKA (red filled circles) are shown in Fig.~\ref{fig:tot_eff} as a function of total particle kinetic energy per nucleon.

\begin{figure}[!htb]
	\centering
	\includegraphics[scale=0.5]{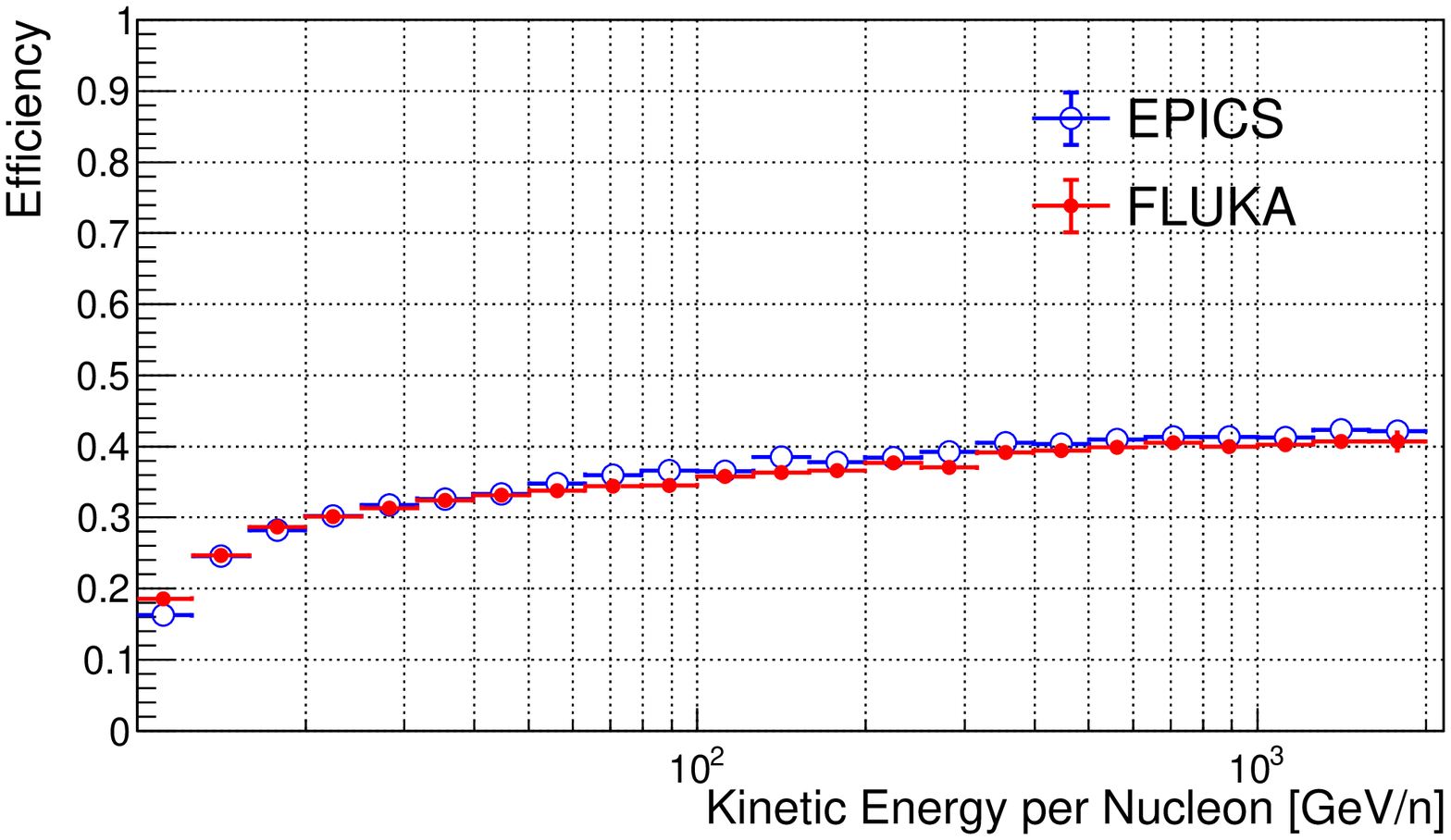}
	\caption{Total selection efficiency for iron events as estimated with EPICS (blue open circles) and FLUKA (red filled circles) simulations.
	}
	\label{fig:tot_eff}
\end{figure}
The above efficiencies were validated by comparing distributions relevant to the selection of events, and obtained from flight data, with the same distributions generated by EPICS or FLUKA. 

\begin{figure}[!htb] 
	\centering
	\includegraphics[scale=0.7]{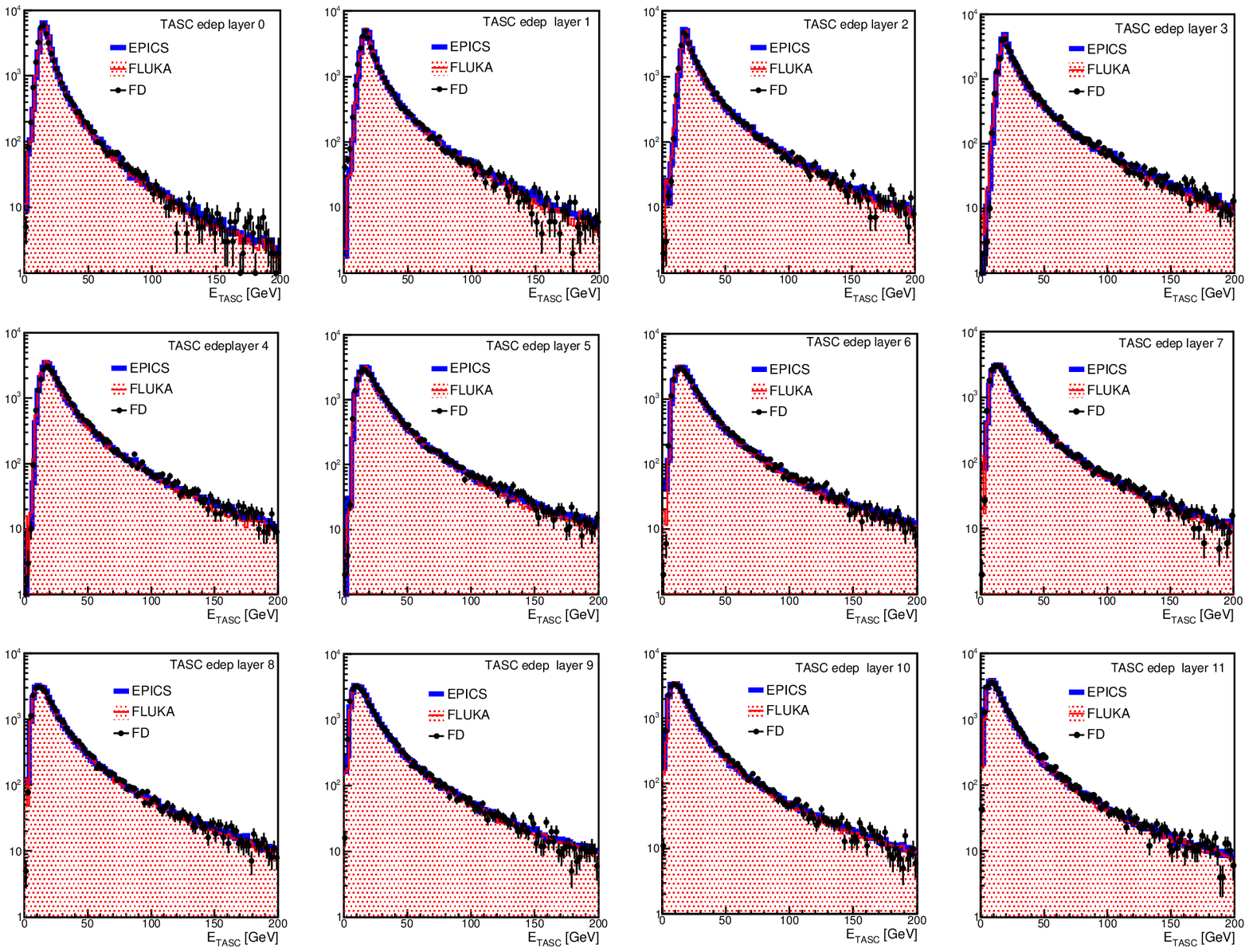}
	\caption{
		Energy observed in each of the 12 layers of the TASC (black points) for the final sample of iron candidates from flight data. It is compared with pure samples of iron simulated by EPICS (blue) or FLUKA (red). 
	}
	\label{fig:TASC_layers}
\end{figure}

An example is given in Fig.~\ref{fig:TASC_layers} where the total energy observed in each layer of the TASC (black points) were plotted using the final sample of iron candidates, marginally contaminated by a residual background (estimated around $1\%$) due to elements with atomic number close to iron. Compared with pure iron samples simulated by EPICS (blue) or FLUKA (red), the distributions from the two MC were found to be consistent with each other and in fair agreement with flight data.
\newline
	
\noindent{\bf{Interactions in the instrument.}} The amount of instrument material above the CHD is very small and well known. The largest significant contribution is limited to a 2 mm thick Al cover placed on top of the CHD. This amounts to $\sim2.2\%$ radiation length and $5\times10^{-3}\lambda_{I} $. The material description in the MC is very accurate and derived directly from the CAD model.  As CALET is sitting on the JEM external platform of the ISS, no extra material external to CALET is normally present within the acceptance adopted for the flux measurement. However, occasional obstructions caused by the ISS robotic arm operations may temporarily affect the FOV. Those rare periods are well identified and events discarded accordingly (a continuous monitoring is routinely done for gamma-ray analyses).   

MC simulations were used to evaluate the iron survival probability after traversing both layers of the CHD and the material above. In order to check its consistency with flight data, the ratio R = (CHDX \& CHDY) $ / $ CHDX (i.e. the fraction of iron candidates tagged by both CHD layers among those detected by the top charge detector) was plotted, as a function of the TASC energy, for selected iron candidates with measured charge in the range 25.5 - 26.5. 
In the upper panel of Fig.~\ref{fig:survival_prob}, R is shown in 15 bins of the TASC energy for both MC ($R_{MC}$) and FD ($R_{FD}$) with an average value around $90\%$ and a flat energy dependence. 
The $R_{MC} / R_{FD} \, \it{double \, ratio}$  (lower plot) shows a good level of consistency between the MC and flight data, within the errors.
The total loss ($\sim 10\%$) of iron events interacting in the upper part of the instrument was taken into account in the total efficiency and its uncertainty included in the systematic error.
\begin{figure}[!htb] 
	\centering
	\includegraphics[scale=0.57]{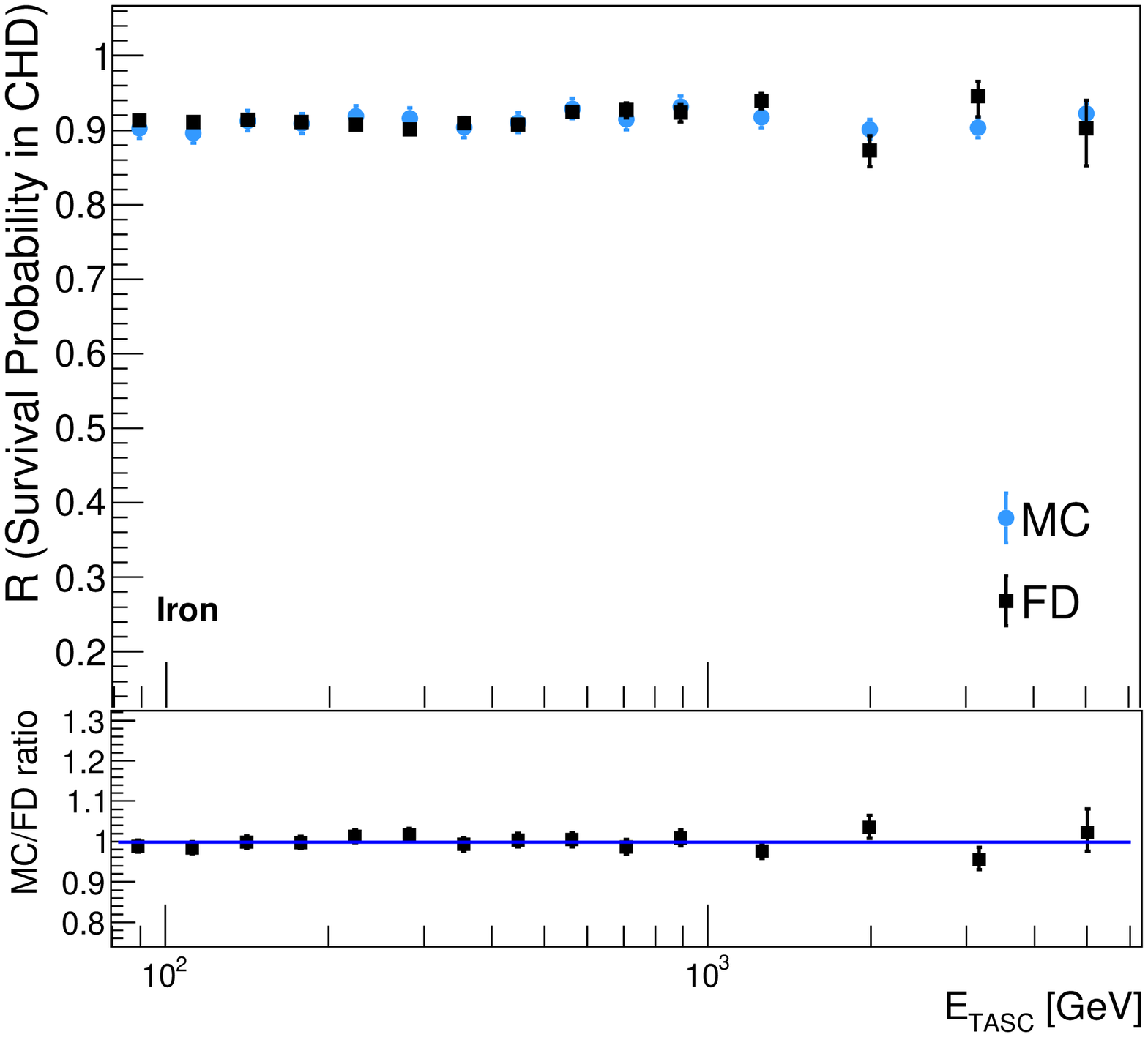}
	\caption{Top panel: iron survival probability R as a function of energy for events crossing both layers of the CHD with flight data (black filled squares) and EPICS (blue filled circles); Bottom panel: ratio of $R_{MC} / R_{FD}$ with the MC and flight data, respectively, fitted with a constant and consistent with unity within the error.
	}
	\label{fig:survival_prob}
\end{figure}
\newline

\noindent{\bf{Background contamination.}}
Background contamination of iron from neighbor elements is relatively small as shown in Fig.~\ref{fig:TASCedepC} as a function of the TASC energy. It is subtracted from the flux as explained in the main body of the paper.
\newline
\begin{figure}[!htb]
	\begin{center}
		\includegraphics[scale=0.6]{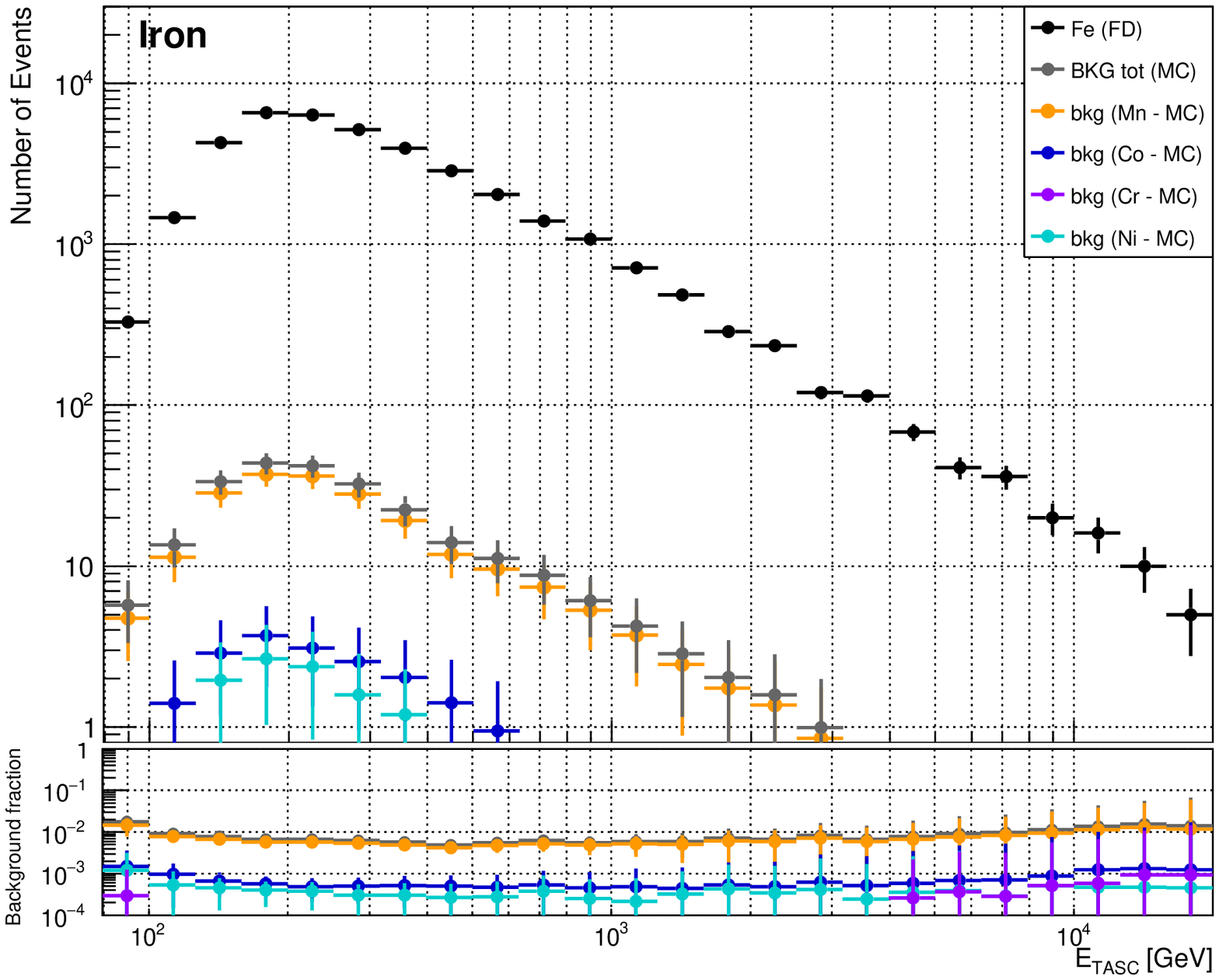}
		\caption{Top panel: Differential distributions of the number of events in a given bin of calorimetric energy ($E_{\rm TASC}$ in GeV) for selected iron events in flight data (black dots) before the unfolding procedure and with background events from nuclei close to iron in atomic number.
			Bottom panel: Contamination from each nuclear species between Z = 24 and Z~=~28 from the MC. The Monte Carlo events are weighted with a factor to reproduce a single power law spectrum with spectral index -2.6, and event selection is the same as for flight data. The resulting elemental charge distribution in each observed energy bin has been normalized to match the CHD charge distribution of flight data. The number of contaminant events is calculated by integration of all the MC events accepted by the iron charge selection.
		}
		\label{fig:TASCedepC}
	\end{center}
\end{figure}
	
\noindent{\bf{Calorimetric energy, bin size, and  unfolding.}}
The energy response of the TASC was studied with MC simulations and compared with the results of measurements of the total particle energy vs beam momentum carried out at CERN.  During one of the beam test campaigns of CALET at the SPS with an extracted primary beam of $^{40}$Ar (150 GeV/c/n), beam fragments were generated from an internal target and guided toward the instrument along a magnetic beam spectrometer that provided an accurate selection of their rigidity and A/Z ratio. The relation between the observed TASC energy and the primary energy was measured for several nuclei up to the highest available energy (6 TeV total particle energy in the case of $^{40}$Ar). After an offline rejection of a very small amount of beam contaminants from the data, the shape of the TASC total energy was found to be consistent with a Gaussian distribution~\cite{akaike2015-SM}. 

The correlation matrix used for the unfolding was derived from the simulations, using two different MC codes EPICS and FLUKA, and applying the same selection cuts as in the FD analysis. Three different binning schemes (4, 5, 10 equal log-bins /decade) have been applied.  Two normalized unfolding matrices (with 4 and 10 bins/decade) derived from EPICS are shown in Fig.~\ref{fig:UFmatrix}.

\begin{figure}[!htb]
	\centering
	\subfigure{	\includegraphics[width=0.45\hsize]{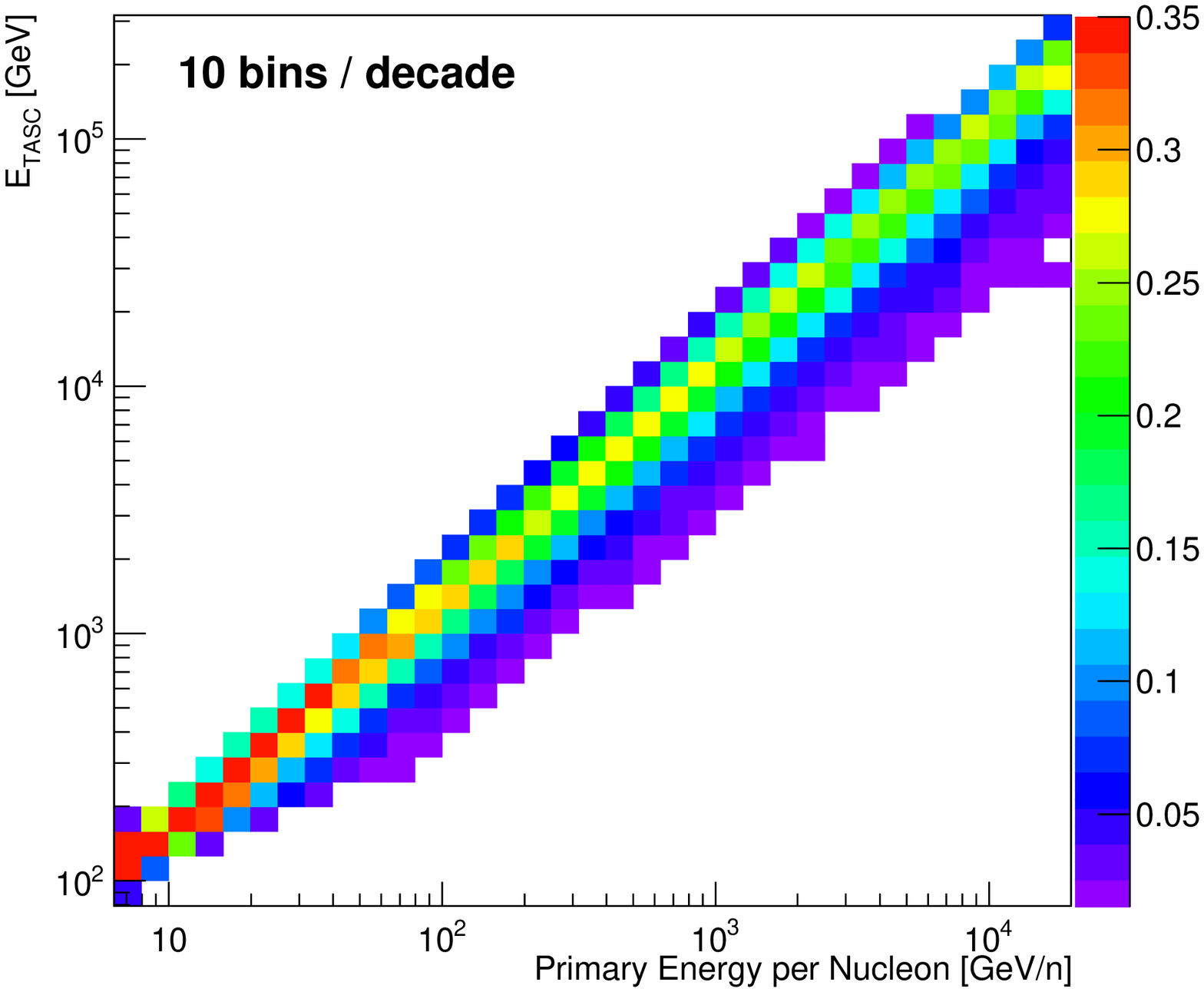}}
	\subfigure{\includegraphics[width = 0.45\hsize]{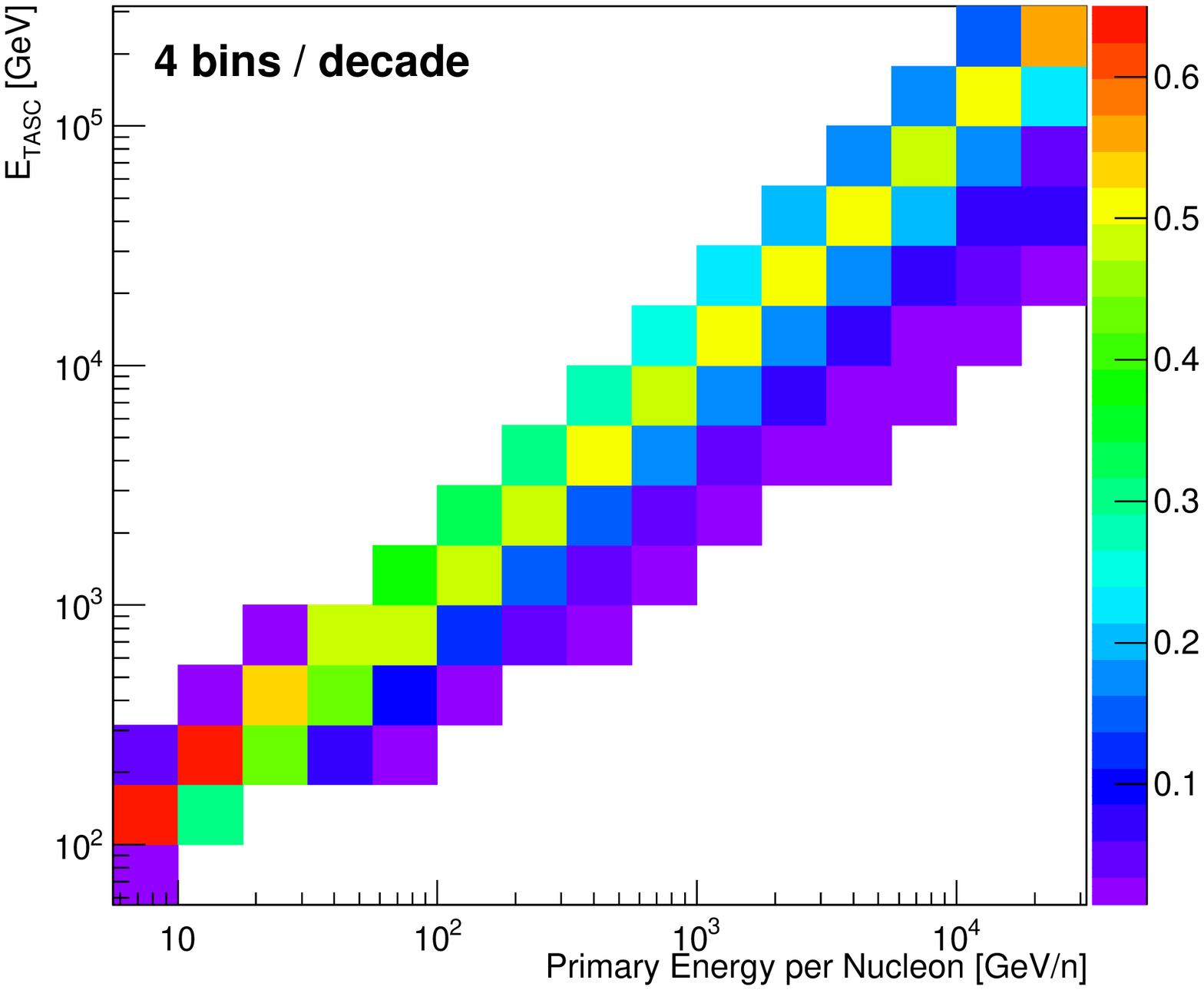}}
	\caption{Response matrix for iron derived from the MC simulations of the CALET flight model by applying the same selection as for flight data. The array is normalized so that the color scale is associated to the probability that iron candidates in a given bin of particle kinetic energy cover different intervals of $ E_\mathrm{TASC} $\label{fig:UFmatrix}. Left: 10 bins/decade; Right: 4 bins/decade.
	} 
\end{figure}

The aforementioned three binning schemes were applied to the iron flux analysis with the result shown in 
Fig.~\ref{fig:flux_with_3_binnings} where only statistical errors are shown. Within the errors, no statistically significant difference was found among the three fluxes.
\newline
\begin{figure}[!htb]
	\centering
	\includegraphics[width=0.7\hsize]{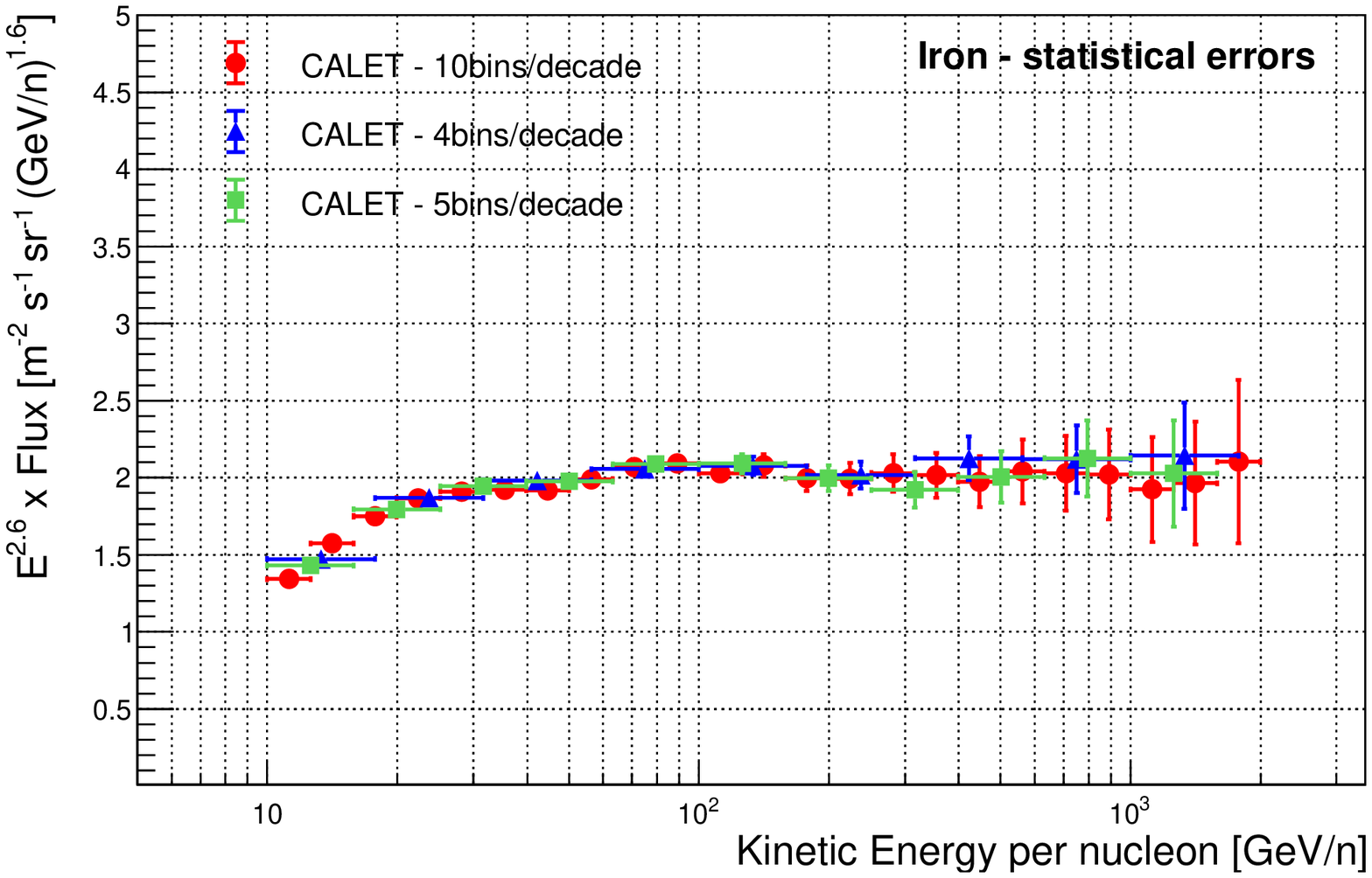}
	\caption{CALET iron flux with 10 bins/decade (red circles), 5 bins/decade (green squares) and 4 bins/decade (blue triangles). The error bars are representative of purely statistical errors.
	}
	\label{fig:flux_with_3_binnings}
\end{figure}
	
\noindent{\bf{Energy dependent systematic errors.}}
A breakdown of energy dependent systematic errors stemming from several sources (as explained in the main body of the paper) and including selection cuts, charge identification, MC model, energy scale correction, energy unfolding, beam test configuration and shower event shape is shown in Fig.~\ref{fig:sys_all} as a function of kinetic energy per nucleon. 
\newline
\begin{figure}[!hbt]
	\centering
	\includegraphics[scale=0.8]{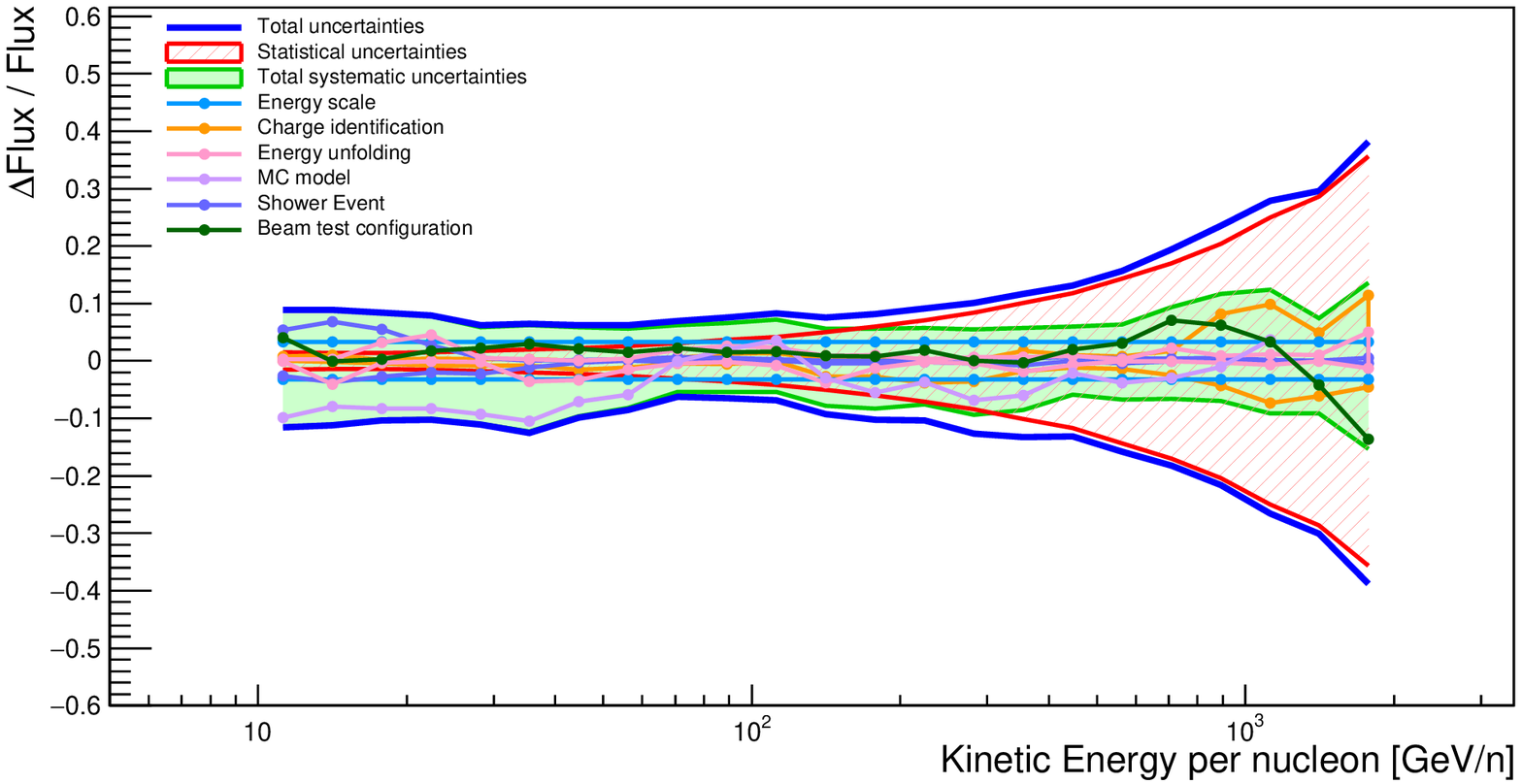}                          
	\caption{Energy dependence (in GeV/nucleon) of systematic uncertainties (relative errors) for iron.  The band bounded by the red lines represents the statistical error. The shaded band within the green lines shows the sum in quadrature of all the sources of systematics including energy independent ones.  A detailed breakdown of systematic energy dependent errors, stemming from charge identification, MC model, energy scale correction, energy unfolding, beam test configuration and shower event shape is shown. The blue lines represent the sum in quadrature of statistical and total systematic uncertainties.
	}
	\label{fig:sys_all}
\end{figure}

\noindent{\bf{Iron flux normalization and spectral shape.}}
The CALET iron flux and a compilation of available data, including the recent AMS-02 measurement~\cite{AMS-Fe-SM}, are shown in Fig.~\ref{fig:FluxSM}, as an enlarged version of Fig.~2 in the main body of the paper.

\begin{figure}[!htb]
	\centering
	\includegraphics[width=0.9\hsize]{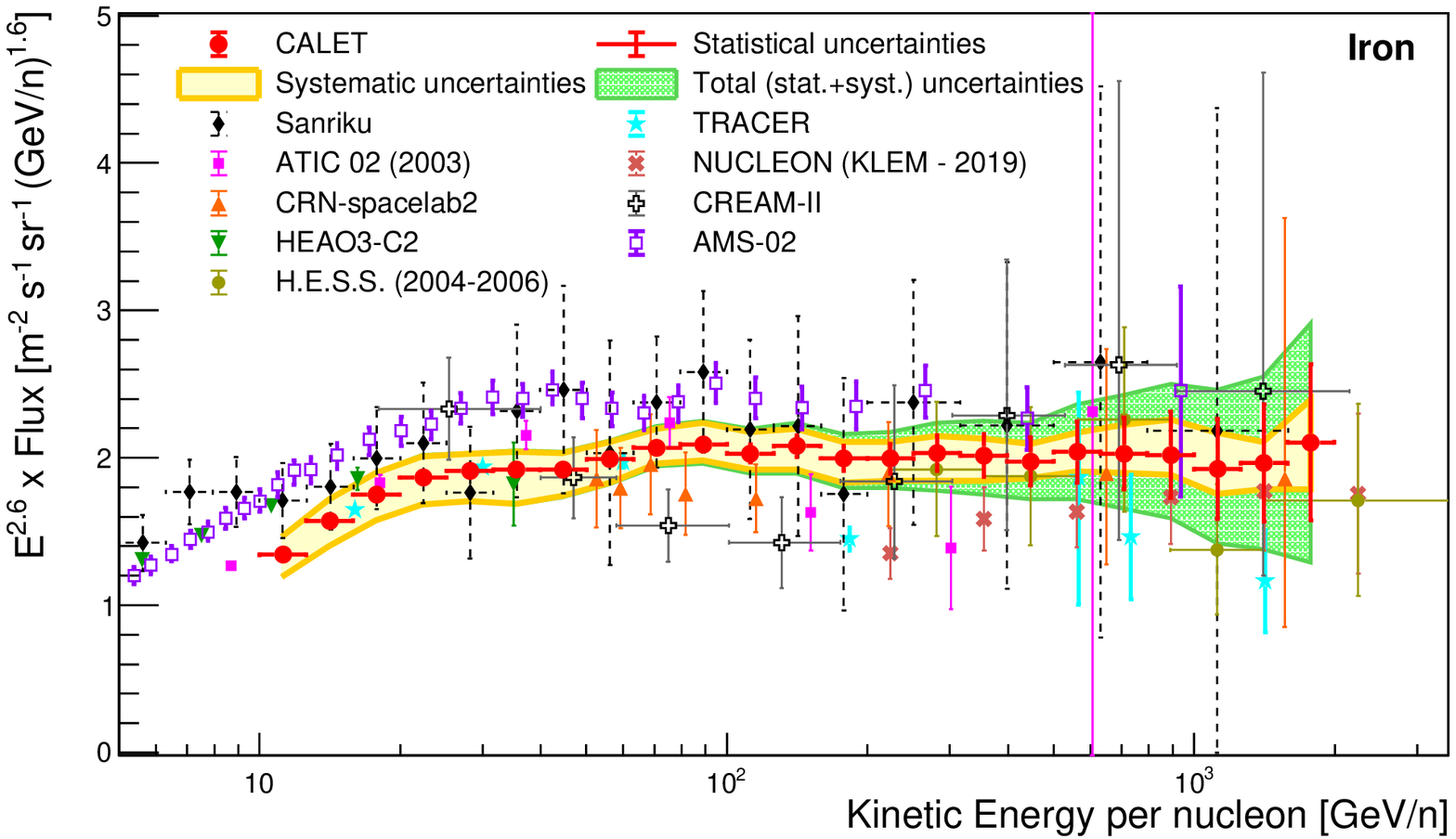}
	\caption{CALET iron flux as a function of kinetic energy per nucleon in GeV (with multiplicative factor $E^{2.6}$). The error bars of the CALET data (red filled circles) represent the statistical uncertainty only. The yellow band indicates the quadrature sum of systematic errors, while the green band indicates the quadrature sum of statistical and systematic errors. Also plotted are the data points from other direct measurements \cite{AMS-Fe-SM, ATIC2-SM, TRACER2008-SM, CREAM2-SM, NUCLEON2019-SM, pallone-SM,HESS-SM,CRN-SM,HEAO-SM}. \label{fig:FluxSM}}
\end{figure}
Normalization issues among cosmic-ray flux measurements have a long historical record and unfortunately they seem to  persist, in a few cases, also among more recent precision measurements. If we focus on the last 15 years, we notice: (1) an inconsistency of the DAMPE electron flux with AMS-02 in the energy interval 
from $\sim50$ GeV  to 1 TeV, whereby the DAMPE electron data are found to be significantly higher than AMS-02 and CALET, the latter being in excellent agreement with each other;  (2) an inconsistency of the DAMPE proton flux with AMS-02 (interval from $\sim300$ GeV to $\sim1$ TeV) and with CALET (from 300 GeV to $\sim$ 10 TeV) where the DAMPE proton flux is significantly higher than the other two experiments;  (3) a tension in the AMS-02 normalization (by more than $20\%$) for carbon flux with respect to PAMELA, CALET and other previous experiments; a similar problem for oxygen with CALET (no oxygen data are available from PAMELA), while preliminary C, O results from DAMPE appear to be consistent in normalization with those of AMS-02  (the spectral shape is consistent within the errors for all three experiments); (4) a tension in normalization (by about $20\%$) between AMS-02 and CALET iron data.

A thorough investigation on possible unaccounted systematic effects related to the normalization of the fluxes has been carried out in the framework of different analyses of CALET with electron, proton, helium, carbon, oxygen, heavier nuclei, and iron data.  We partially or totally rule out specific sources of uncertainty. The presence of a significant systematic issue on the live time normalization of the flux is considered unlikely, given the consistency in normalization of the CALET electron, (as well as proton) flux with AMS-02 (and PAMELA) within their respective rigidity ranges. Possible areas where systematic effects could be further investigated by CALET, and by other collaborations as well, include the impact of simulations (specifically on the hadronic cross sections). While EPICS and FLUKA differences were taken into account in the assessment of systematic errors, a comparative study with GEANT4 is not available for CALET at present. Another area under scrutiny is the CALET trigger efficiency. It was extensively studied using ratios of different trigger modes vs the minimum bias trigger and an excellent agreement was found with the MC simulations (see~\cite{CALET-CO-SM}). The possibility of implementing additional trigger types, and dedicated orbital run modes, to further investigate this aspect is under study.

In figure Fig.~\ref{fig:Comparison-with-AMS02-iron} the recently published AMS-02 flux is compared with the CALET iron data (with the same multiplicative flux factor $ E^{2.7} $), after increasing the latter's overall normalization by $20\%$.
Taking the data at face-value, we notice that the data points of the two experiments not only share a very similar spectral shape, but also have comparable errors. 
	
\begin{figure}[!htb]
	\centering
	\includegraphics[scale=0.7]{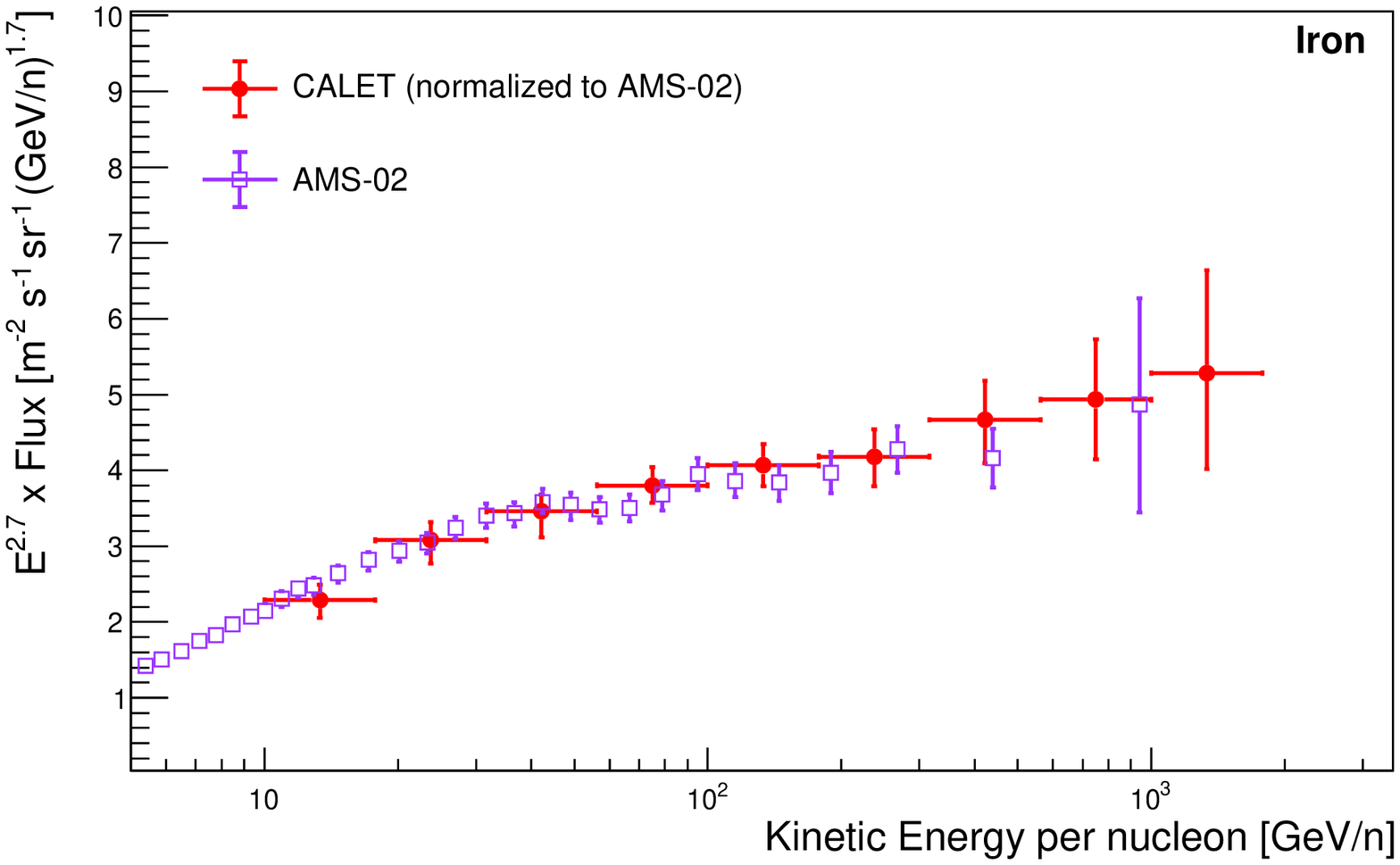}
	\caption{Iron flux (with multiplicative factor $ E^{2.7} $) measured by CALET (red points) with 4 bins/decade, multiplied by 1.20 for comparison with the AMS-02 results~\cite{AMS-Fe-SM}. The error bars of the CALET data are the quadrature sum of statistical and systematic uncertainties.
	}
	\label{fig:Comparison-with-AMS02-iron}
\end{figure}

\clearpage
\renewcommand{\arraystretch}{1.25}
\begin{table*}
	\caption{Table of the CALET differential spectrum in kinetic energy per nucleon of cosmic-ray iron. 
		The first, second, and third error in the flux are representative of the statistical uncertainties, systematic uncertainties in normalization, and energy dependent systematic uncertainties, respectively.
		\label{tab:Cflux}}
	\begin{ruledtabular}
		\begin{tabular}{c c c c}
			Energy Bin [GeV$/n$] & Flux [m$^{-2}$sr$^{-1}$s$^{-1}$(GeV$/n$)$^{-1}$]   \\
			\hline
			10.0 -- 12.6 & $( 2.50  \, \pm 0.03  \, _{- 0.13 }^{+ 0.14 }  \, _{- 0.25 }^{+ 0.17 }) \times 10^{- 3 }$ \\
			12.6 -- 15.8 & $( 1.61  \, \pm 0.02  \, _{- 0.09 }^{+ 0.09 }  \, _{- 0.16 }^{+ 0.11 }) \times 10^{- 3 }$ \\
			15.8 -- 20.0 & $( 9.84  \, \pm 0.10  \, _{- 0.53 }^{+ 0.53 }  \, _{- 0.85 }^{+ 0.62 }) \times 10^{- 4 }$ \\
			20.0 -- 25.1 & $( 5.77  \, \pm 0.06  \, _{- 0.31 }^{+ 0.31 }  \, _{- 0.50 }^{+ 0.32 }) \times 10^{- 4 }$ \\
			25.1 -- 31.6 & $( 3.25  \, \pm 0.04  \, _{- 0.17 }^{+ 0.18 }  \, _{- 0.31 }^{+ 0.08 }) \times 10^{- 4 }$ \\
			31.6 -- 39.8 & $( 1.79  \, \pm 0.03  \, _{- 0.10 }^{+ 0.10 }  \, _{- 0.20 }^{+ 0.05 }) \times 10^{- 4 }$ \\
			39.8 -- 50.1 & $( 9.84  \, \pm 0.16  \, _{- 0.53 }^{+ 0.53 }  \, _{- 0.79 }^{+ 0.21 }) \times 10^{- 5 }$ \\
			50.1 -- 63.1 & $( 5.61  \, \pm 0.11  \, _{- 0.30 }^{+ 0.30 }  \, _{- 0.35 }^{+ 0.09 }) \times 10^{- 5 }$ \\
			63.1 -- 79.4 & $( 3.21  \, \pm 0.07  \, _{- 0.17 }^{+ 0.17 }  \, _{- 0.02 }^{+ 0.10 }) \times 10^{- 5 }$ \\
			79.4 -- 100.0 & $( 1.78  \, \pm 0.05  \, _{- 0.10 }^{+ 0.10 }  \, _{- 0.01 }^{+ 0.07 }) \times 10^{- 5 }$ \\
			100.0 -- 125.9 & $( 9.49  \, \pm 0.28  \, _{- 0.51 }^{+ 0.51 }  \, _{- 0.08 }^{+ 0.45 }) \times 10^{- 6 }$ \\
			125.9 -- 158.5 & $( 5.35  \, \pm 0.19  \, _{- 0.29 }^{+ 0.29 }  \, _{- 0.30 }^{+ 0.07 }) \times 10^{- 6 }$ \\
			158.5 -- 199.5 & $( 2.82  \, \pm 0.12  \, _{- 0.15 }^{+ 0.15 }  \, _{- 0.18 }^{+ 0.04 }) \times 10^{- 6 }$ \\
			199.5 -- 251.2 & $( 1.55  \, \pm 0.08  \, _{- 0.08 }^{+ 0.08 }  \, _{- 0.08 }^{+ 0.03 }) \times 10^{- 6 }$ \\
			251.2 -- 316.2 & $( 8.66  \, \pm 0.52  \, _{- 0.46 }^{+ 0.47 }  \, _{- 0.66 }^{+ 0.06 }) \times 10^{- 7 }$ \\
			316.2 -- 398.1 & $( 4.72  \, \pm 0.34  \, _{- 0.25 }^{+ 0.26 }  \, _{- 0.31 }^{+ 0.09 }) \times 10^{- 7 }$ \\
			398.1 -- 501.2 & $( 2.54  \, \pm 0.21  \, _{- 0.14 }^{+ 0.14 }  \, _{- 0.06 }^{+ 0.06 }) \times 10^{- 7 }$ \\
			501.2 -- 631.0 & $( 1.44  \, \pm 0.15  \, _{- 0.08 }^{+ 0.08 }  \, _{- 0.06 }^{+ 0.05 }) \times 10^{- 7 }$ \\
			631.0 -- 794.3 & $( 7.89  \, \pm 0.95  \, _{- 0.42 }^{+ 0.43 }  \, _{- 0.31 }^{+ 0.60 }) \times 10^{- 8 }$ \\
			794.3 -- 1000.0 & $( 4.32  \, \pm 0.62  \, _{- 0.23 }^{+ 0.23 }  \, _{- 0.19 }^{+ 0.45 }) \times 10^{- 8 }$ \\
			1000.0 -- 1258.9 & $( 2.26  \, \pm 0.40  \, _{- 0.12 }^{+ 0.12 }  \, _{- 0.17 }^{+ 0.25 }) \times 10^{- 8 }$ \\
			1258.9 -- 1584.9 & $( 1.27  \, \pm 0.26  \, _{- 0.07 }^{+ 0.07 }  \, _{- 0.09 }^{+ 0.06 }) \times 10^{- 8 }$ \\
			1584.9 -- 1995.3 & $( 7.47  \, \pm 1.88  \, _{- 0.40 }^{+ 0.40 }  \, _{- 1.08 }^{+ 0.93 }) \times 10^{- 9 }$ \\
		\end{tabular}
	\end{ruledtabular}
\end{table*}
\renewcommand{\arraystretch}{1.0}
	\providecommand{\noopsort}[1]{}\providecommand{\singleletter}[1]{#1}%
%

\end{document}